\documentclass[fleqn,10pt]{wlscirep}
\usepackage[utf8]{inputenc}
\usepackage[T1]{fontenc}

\usepackage{xr}

\RenewDocumentEnvironment{abstract}{+b}{%
    \gdef\theabstract{\parbox{\linewidth}{\internallinenumbers #1}}%
}{}

\usepackage{siunitx}
\usepackage{nicefrac}
\usepackage{setspace}
\usepackage{gensymb}
\usepackage{xcolor}
\usepackage{amsmath}
\usepackage{ragged2e}
\usepackage{hyperref}
\usepackage{cleveref}
\usepackage[mathlines,running]{lineno}
\usepackage{url}

\DeclareMathAlphabet\mathbfcal{OMS}{cmsy}{b}{n}
\DeclareSIUnit\angstrom{\text {Å}}
\DeclareUnicodeCharacter{03B4}{$\delta$}
\DeclareUnicodeCharacter{2212}{\textendash}
\DeclareMathOperator\erf{erf}

\title{Ultrafast decoupling of polarization and strain in ferroelectric $\mathrm{BaTiO_3}$}

\author[1,2,3]{Le Phuong Hoang}
\author[4]{David Pesquera}
\author[5]{Gerard N. Hinsley}
\author[1]{Robert Carley}
\author[1]{Laurent Mercadier}
\author[1]{Martin Teichmann}
\author[4]{Saptam Ganguly}
\author[1]{Teguh Citra Asmara}
\author[6,1]{Giacomo Merzoni}
\author[1]{Sergii Parchenko}
\author[1]{Justine Schlappa}
\author[1]{Zhong Yin}
\author[4]{José Manuel Caicedo Roque}
\author[4]{José Santiso}
\author[7]{Irena Spasojevic}
\author[1]{Cammille Carinan}
\author[8]{Tien-Lin Lee}
\author[3,9]{Kai Rossnagel}
\author[8]{Jörg Zegenhagen}
\author[4,10]{Gustau Catalan}
\author[5]{Ivan A. Vartanyants}
\author[1]{Andreas Scherz}
\author[1,*]{Giuseppe Mercurio}

\affil[*]{giuseppe.mercurio@xfel.eu}
\affil[1]{European XFEL, 22869 Schenefeld, Germany}
\affil[2]{Max Planck Institute for the Structure and Dynamics of Matter, 22761 Hamburg, Germany}
\affil[3]{Institute of Experimental and Applied Physics, Kiel University, 24098 Kiel, Germany}
\affil[4]{Catalan Institute of Nanoscience and Nanotechnology (ICN2), CSIC and BIST, Campus UAB, Bellaterra, 08193 Barcelona, Spain}
\affil[5]{Deutsches Elektronen-Synchrotron DESY, 22607 Hamburg, Germany}
\affil[6]{Dipartimento di Fisica, Politecnico di Milano, 20133, Milano, Italy}
\affil[7]{Department de Física, Universitat Autònoma de Barcelona, 08193 Bellaterra, Spain}
\affil[8]{Diamond Light Source Ltd., Didcot, OX110DE Oxfordshire, United Kingdom}
\affil[9]{Ruprecht Haensel Laboratory, Deutsches Elektronen-Synchrotron DESY, 22607 Hamburg, Germany}
\affil[10]{Institucio Catalana de Recerca i Estudis Avançats (ICREA), 08010 Barcelona, Catalonia}


\clearpage

\begin{abstract}
A fundamental understanding of the interplay between lattice structure, polarization and electrons is pivotal to the optical control of ferroelectrics. The interaction between light and matter enables the remote and wireless control of the ferroelectric polarization on the picosecond timescale, while inducing strain, i.e., lattice deformation. At equilibrium, the ferroelectric polarization is proportional to the strain, and is typically assumed to be so also out of equilibrium. Decoupling the polarization from the strain would remove the constraint of sample design and provide an effective knob to manipulate the polarization by light. Here, upon an above-bandgap laser excitation of the prototypical ferroelectric BaTiO$_3$, we induce and measure an ultrafast decoupling between polarization and strain that begins within \SI{350}{\fs}, by softening Ti-O bonds via charge transfer, and lasts for several tens of picoseconds. We show that the ferroelectric polarization out of equilibrium is mainly determined by photoexcited electrons, instead of the strain. This excited state could serve as a starting point to achieve stable and reversible polarization switching via THz light. Our results demonstrate a light-induced transient and reversible control of the ferroelectric polarization and offer a pathway to control by light both electric and magnetic degrees of freedom in multiferroics.
\end{abstract}

\begin{document}


\flushbottom
\maketitle

\clearpage

Ferroelectric materials are characterized by many properties, including piezoelectricity and pyroelectricity, besides ferroelectricity, which make them attractive for a wide range of applications, such as nonvolatile memories, transistors, sensors, and actuators\cite{martin_thin-film_2017,kim_ferroelectric_2023}. The key property of a ferroelectric material is the ability to switch its spontaneous polarization in response to an external electric field. This is typically achieved by a static or pulsed electric field with the consequent limitations given by the need for complex circuitry and switching times of hundreds of picoseconds to nanoseconds\cite{guo_recent_2021}. These challenges can be overcome by optical control of the ferroelectric polarization. Light-matter interaction enables remote and wireless control of the ferroelectric polarization on the picosecond timescale\cite{guo_recent_2021}. Moreover, since all ferroelectrics are also piezoelectrics, the ferroelectric polarization is strongly coupled to the strain, i.e., the lattice deformation\cite{ruello_physical_2015}. Optical control of polarization and strain has been achieved in several cases. For example, in multilayers of ferroelectric and electrode thin films, an optical laser was used to excite the metal (or semiconductor) layer and indirectly the ferroelectric material, leading to a transient modification of the strain\cite{korff_schmising_coupled_2007, schick_ultrafast_2014, sheu_using_2014, lee_structural_2021}, or the polarization by charge redistribution at the interface\cite{li_optical_2018}. In other studies, light was absorbed directly by the ferroelectric material, inducing changes in the spontaneous polarization\cite{ron_transforming_2023, sarott_reversible_2024} or lattice strain in clamped\cite{daranciang_ultrafast_2012, wen_electronic_2013, schick_localized_2014, matzen_tuning_2019, ahn_dynamic_2021, sri_gyan_optically_2022, gu_temporal_2023} or freestanding\cite{ganguly_photostrictive_2024} ferroelectric thin films. THz light was employed to rotate\cite{chen_ultrafast_2016} or even transiently reverse the orientation of the spontaneous polarization \cite{mankowsky_ultrafast_2017}. In all these studies so far, either the polarization or the strain was measured, and a direct proportionality between spontaneous polarization and strain was typically assumed\cite{yang_photostriction_2018}. This proportionality is based on the piezoelectric effect, which is well captured by the Landau-Ginzburg-Devonshire theory when the lattice distortion is along the polarization axis\cite{devonshire_xcvi_1949}. While this assumption is valid under equilibrium conditions, as demonstrated experimentally, e.g. in Refs.\cite{choi_enhancement_2004, dawber_tailoring_2007}, it may not hold under out-of-equilibrium conditions following light-matter interaction. Decoupling the polarization from the strain would remove the constraint of sample design to achieve specific properties \cite{choi_enhancement_2004, pesquera_beyond_2020}, and, at the same time, would provide a more effective and ultrafast knob to manipulate the polarization by light. 

To explore this scenario, we probe the out-of-plane strain and the spontaneous polarization of the prototypical ferroelectric BaTiO$_3$ upon above-bandgap absorption of ultrashort UV light pulses (Figure \ref{fig:Figure_1}a). A fundamental understanding of the relationship between strain and ferroelectric polarization out of equilibrium requires their investigation on their natural timescale encompassing $\approx$ \SI{100}{\fs} to several tens of \SI{}{\ps}. We employ a combination of time-resolved X-ray diffraction (tr-XRD), time-resolved optical second harmonic generation (tr-SHG), and time-resolved optical reflectivity (tr-refl) to obtain the magnitudes and the separate dynamics of the out-of-plane lattice parameter, the spontaneous polarization, and the photoexcited carrier density, respectively\cite{zhang_probing_2021}, with a time resolution of $\approx \SI{90}{\fs}$. In this paper, we will show the mechanisms that govern the structure and polarization changes in a ferroelectric material and their complex relationship out of equilibrium in the presence of photoexcited electrons and lattice deformation. In particular, since the strain wave propagates at the speed of sound, whereas electronic interactions are much faster, we induce and measure an ultrafast decoupling between polarization and strain, which we assign to the photoexcited electrons. First, we present the lattice response to the absorption of UV laser pulses and the corresponding data modeling. Next, we present tr-SHG and tr-refl data. Finally, we bring together all the results and discuss the underlying physical mechanisms in the context of hitherto known phenomena taking place in ferroelectric materials.

\clearpage

\section*{Results}
\textbf{Photoinduced structural dynamics}

Our sample consists of a coherently strained, monodomain BaTiO$_3$ (BTO) thin film, grown on a GdScO$_3$ (GSO) substrate, with a SrRuO$_3$ (SRO) bottom electrode sandwiched in between (see Methods). Under a compressive strain of $-0.55\%$ imposed by the substrate, the BTO film shows an out-of-plane ferroelectric polarization $P_s$ pointing toward the sample surface (Figure \ref{fig:Figure_1}a). The sample is excited above the BTO band gap $E_g = \SI{3.4}{\eV}$\cite{chernova_strain-controlled_2015} using \SI{266}{\nm} laser pulses at an incident pump laser fluence of \SI{2.7}{\mJ\per\square\cm}. Time-resolved X-ray diffraction of the (001) Bragg reflection is employed to probe the lattice response of our ferroelectric thin film along the out-of-plane direction. The lattice deformations along the in-plane directions on the picosecond timescale are negligible, given the large ratio between photoexcited area ($330\times240$ \SI{}{\micro\metre\squared}) and BTO film thickness $d_\mathrm{BTO} = \SI{34.5}{\nm}$.

We observe an initial reduction of the tetragonal distortion, which goes hand in hand with lattice compression, then followed by lattice expansion. In particular, Figure \ref{fig:Figure_1}b shows the (001) diffraction intensity $I_\mathrm{XRD}$ of BTO as a function of the photon energy $E_\nu$ and at different pump-probe delays $t$ from $\SI{-14}{\ps}$ to $\SI{32.5}{\ps}$. At $t = \SI{2.6}{\ps}$ we observe the following changes to the Bragg peak as compared to the ground state (at $t = \SI{-14}{\ps}$): a decrease in the diffraction intensity $I_\mathrm{XRD}$ near the peak center, and a shift to higher photon energy, which implies a decrease in the out-of-plane lattice parameter $c$, i.e., lattice compression (see gray curve in Figure \ref{fig:Figure_1}b). To further explore this initial structural dynamics, we measure the delay dependence of $\Delta I_\mathrm{XRD} / I_\mathrm{XRD,0} = [I_\mathrm{XRD}(t) - I_\mathrm{XRD,0}] / I_\mathrm{XRD,0}$, which quantifies the relative change of $I_\mathrm{XRD}(t)$ at the photon energy $E_\nu=\SI{1525}{\eV}$ of the BTO peak with respect to the equilibrium value $I_\mathrm{XRD,0}$ at negative delays. We observe a maximum diffraction intensity drop of $\approx 4\%$ at $t = \SI{3.5}{\ps}$, with up to $99\%$ recovery to the equilibrium value at $t\approx\SI{7}{\ps}$ (Figure \ref{fig:Figure_1}c and \ref{si:fig:diffraction_curve_drop}). We assign the initial $4\%$ drop and recovery in diffraction intensity to the displacements of atoms within the BTO unit cell (inset of Figure \ref{fig:Figure_1}c). Specifically, simulations based on the dynamical theory of diffraction (\ref{si:fig:DW_effect}) exclude the Debye-Waller effect and show that a decrease in the displacement $\Delta_\mathrm{Ti-O}$ between the Ti atom and the center of the O octahedron by \SI{8}{\pm} can model the measured maximum change in peak diffraction intensity.

We focus next on the BTO (001) Bragg peak measured at longer time delays (Figure \ref{fig:Figure_1}b). We observe that $I_\mathrm{XRD}(E_\nu)$ at $t>\SI{3}{\ps}$ are shifted toward lower photon energies, i.e., larger out-of-plane lattice parameters $c$, with respect to $I_\mathrm{XRD}(E_\nu)$ at smaller delays $t$. This can be clearly seen from the plot of the BTO out-of-plane strain $\overline{\eta}(t)$, averaged over $d_\mathrm{BTO}$ (Figure \ref{fig:Figure_1}d). Here, $\overline{\eta}(t) = [c(t) - c_0]/c_0$, with $c(t)$ and $c_0$ representing the average $\overline{c}$ at a given $t > \SI{0}{\ps}$ and $t \leq \SI{0}{\ps}$, respectively (see Methods). In Figure \ref{fig:Figure_1}d, we find that: (i) the maximum compression of $-0.024\%$ occurs at $t=\SI{2.6}{\ps}$, (ii) $\overline{\eta}(t)$ increases linearly at a rate of $\SI{0.04}{\%\per\ps}$ in the range $\SI{4}{\ps}<t<\SI{10}{\ps}$, and (iii) settles at $0.34\%$ at $\approx \SI{20}{\ps}$.

The model fitting $\overline{\eta}(t)$ data in Figure \ref{fig:Figure_1}d is presented in the following. When a photon with energy $E > E_g$ is absorbed in BTO, electrons are photoexcited from the O 2p-derived valence band to the Ti 3d-derived conduction band \cite{lian_indirect_2019, chen_ferroelectric_2024, kolezynski_molecular_2005}. The thermalization of photoexcited electrons leads to an increase in the electron temperature ($T_e$), and to changes in the electronic system that can be modeled by the variation of the bandgap as a function of the electronic pressure ($\partial E_g / \partial p$) \cite{ruello_physical_2015}. In turn, a modified electron system affects the interatomic potential, resulting in atomic motions and contributing to the deformation potential stress $\sigma_{DP}(T_e, \partial E_g / \partial p)$. Subsequently, photoexcited electrons transfer part of their excess energy ($E-E_g$) to the phonon system via electron-phonon scattering, increasing the phonon temperature ($T_p$) on the picosecond timescale. This, in turn, induces a lattice expansion dependent on the BTO linear expansion coefficient ($\beta$), and contributes to the thermoelastic stress $\sigma_{TE}(T_p, \beta)$. The total stress\cite{ruello_physical_2015, wright_acoustic_1995} $\sigma = \rho v^2 \eta + \sigma_{DP} + \sigma_{TE}$ generates a strain wave $\eta(z,t)$ that propagates through the material of mass density $\rho$ at the longitudinal speed of sound $\nu$. We solve analytically the two-temperature model (\ref{si:note:2tm}) and the lattice strain wave equation (\ref{si:note:strain_model}) to obtain $\eta(z,t)$. Finally, we calculate the strain $\overline{\eta}(t)$, averaged over $d_\mathrm{BTO}$, to fit the experimental data in Figure \ref{fig:Figure_1}d. The main outcome of our fit model is a negative $\partial E_g / \partial p$ of the order of $\approx\SI{e-31}{\joule\per\pascal}$, in agreement with first-principles calculations\cite{khenata_first-principle_2005}, with a resulting bandgap decrease of about \SI{3.2}{\milli\eV} (\ref{si:note:estimation_bandgap}). The negative $\partial E_g / \partial p$ causes lattice compression within the first $\approx \SI{3}{\ps}$, when $\sigma_{DP}$ dominates over $\sigma_{TE}$ (inset of Figure \ref{fig:Figure_1}d). Conversely, at larger time delays ($t>\SI{3}{\ps}$), $T_p$ increases (\ref{si:fig:Temperature_profile}b) and the thermoelastic term becomes the dominant one, leading to lattice expansion (\ref{si:fig:TE_DP_strain}).

\textbf{Photoinduced ferroelectric polarization and electron dynamics}

We turn now to investigating the dynamics of the ferroelectric polarization $P_s$ and of the photoexcited carriers \cite{chen_ultrafast_2012, jin_structural_2012, sheu_ultrafast_2012, wang_coupling_2019, mudiyanselage_coherent_2019}, upon excitation of the BTO film by the same \SI{266}{\nm} pump laser with fluence \SI{2.7}{\mJ\per\square\cm}. Therefore, we perform tr-SHG experiments \cite{denev_probing_2011, zhang_probing_2021} and simultaneously tr-refl in reflection geometry (Figure \ref{fig:Figure_3_shg_refl}a). From SHG polarimetry, i.e., the dependence of SHG intensity $I_\mathrm{SHG} (\varphi) \propto |\chi^{(2)}_{ijk} E(\omega)^2|^2$ on the polarization angle of the probe beam $\varphi$, we learn about the optical tensor elements $\chi^{(2)}_{ijk}$ of a material, and thus its symmetry\cite{denev_probing_2011}. By selecting either horizontal ($p$) or vertical ($s$) polarization of the SHG beam, we measure $I_\mathrm{SHG}^{p}(\varphi)$ and $I_\mathrm{SHG}^{s}(\varphi)$, shown in Figures \ref{fig:Figure_3_shg_refl}b and c (blue points) together with the respective fit curves (see Methods), which are based on the 4mm point group symmetry with the following nonzero tensor elements: $\chi_{zxx}^{(2)}$, $\chi_{xxz}^{(2)}$, and $\chi_{zzz}^{(2)}$.

We observe a reduction of the BTO tetragonality, upon laser excitation, from the time evolution of $\chi_{zxx}^{(2)}$, $\chi_{xxz}^{(2)}$, and $\chi_{zzz}^{(2)}$ in the delay range $\SI{-1}{\ps} < t < \SI{30}{\ps}$. While the 4mm symmetry is preserved also after the pump excitation (orange points in Figures \ref{fig:Figure_3_shg_refl}b-c), a laser-induced change in tetragonality is observed. To visualize it, we display the relative change $\Delta \chi / \chi_0$ of $\chi_{xxz}^{(2)}$, $\chi_{zxx}^{(2)}$, and $\chi_{zzz}^{(2)}$ as a function of delay $t$ (Figure \ref{fig:Figure_3_shg_refl}d for $\SI{-1}{\ps} < t < \SI{3}{\ps}$ and \ref{si:fig:fit_chi_long_delay_scans} for $\SI{-1}{\ps} < t < \SI{30}{\ps}$). The three tensor elements show similar dynamics, characterized by a fast fall time with the maximum drop after $\approx\SI{500}{\fs}$ and two exponential recovery time constants of $\approx\SI{5.5}{\ps}$ and $\approx\SI{45}{\ps}$. Interestingly, the tensor elements representative of the electric dipole along the out-of-plane direction $z$ ($\chi_{zxx}^{(2)}$ and $\chi_{zzz}^{(2)}$) show a nearly identical time dependence and a larger relative change than $\chi_{xxz}^{(2)}$, which refers to the in-plane electric dipole along the direction $x$. The difference between $\chi_{zxx}^{(2)}$ (or $\chi_{zzz}^{(2)}$) and $\chi_{xxz}^{(2)}$ reaches $0.5\%$ after $\approx\SI{500}{\fs}$ and decreases in a few tens of picoseconds (\ref{si:fig:fit_chi_long_delay_scans}). A purely thermal effect \cite{Murgan_calculation_2002} would cause a uniform change of all tensor elements $\chi^{(2)}_{ijk}$, whereas the measured different dynamics of $\chi^{(2)}_{ijk}$ indicates a time-dependent lattice distortion of non-thermal origin.
In fact, TD-DFT calculations\cite{lian_indirect_2019, chen_ferroelectric_2024} show that upon charge transfer, the Ti-$\mathrm{O_\parallel}$ bonds between Ti and apical $\mathrm{O_\parallel}$ atoms (parallel to $P_s$) are weakened more than Ti-$\mathrm{O_\perp}$ bonds between Ti and basal $\mathrm{O_\perp}$ atoms (perpendicular to $P_s$), with a resulting reduction of the tetragonal distortion (inset of Figure \ref{fig:Figure_3_shg_refl}d). Consequently, it is intuitive to expect a larger amplitude of the induced electric dipole along the Ti-$\mathrm{O_\parallel}$ direction ($z$) with respect to the Ti-$\mathrm{O_\perp}$ direction (in the $xy$ plane), as experimentally demonstrated by our data.

The proportionality $I_\mathrm{SHG} \propto |\chi^{(2)}_{ijk}|^2 \propto |P_s|^2$ gives direct access to the magnitude of the spontaneous polarization \cite{denev_probing_2011}. To this end, we measure the relative change $\Delta I_\mathrm{SHG}^p / I_\mathrm{SHG,0}^p$ as a function of pump-probe delay $t$ (Figure \ref{fig:Figure_3_shg_refl}e), with the polarization of the \SI{800}{\nm} probe beam fixed to the maximum of $I_\mathrm{SHG}^{p}(\varphi)$ at $\varphi = 0\degree$ (Figure \ref{fig:Figure_3_shg_refl}b). Simultaneously, we measure the relative change in reflectivity $\Delta R / R_0$ as a function of pump-probe delay $t$ (Figure \ref{fig:Figure_3_shg_refl}f). The data in Figure \ref{fig:Figure_3_shg_refl}e [f] are well reproduced by a fit function consisting of the sum of three exponential decay terms, with fall [rise] time $\tau_0$, and recovery times $\tau_1$ and $\tau_2$, convoluted with a Gaussian function representing the experimental temporal resolution (\ref{SI:note:fit_function}). The initial drop in SHG intensity by $10 \%$ within $\SI{350}{\fs}$ is followed by $\tau_1^\mathrm{SHG} = \SI{7.2\pm0.5}{\ps}$ and $\tau_2^\mathrm{SHG} = \SI{200\pm140}{\ps}$ recovery times, resulting in a $2.3\%$ drop at $\SI{40}{\ps}$ (Figure \ref{fig:Figure_3_shg_refl}e). At the same time, we observe a fast increase in reflectivity by $7\%$ within $\SI{350}{\fs}$, followed by two recovery times, $\tau_1^{R} = \SI{5.2\pm0.1}{\ps}$ and $\tau_2^{R} = \SI{29.8\pm0.5}{\ps}$ (Figure \ref{fig:Figure_3_shg_refl}f).

In both tr-SHG and tr-refl data, the time needed to reach the maximum relative change ($\SI{350}{\fs}$) might be due to the thermalization of photoexcited electrons via electron-electron scattering. Subsequently, thermalized electrons, which are higher in the conduction band, move to the bottom of the conduction band, transferring energy to the phonon system, and recombining with holes in the valence band via electron-phonon scattering\cite{thomsen_surface_1986, ruello_physical_2015, wang_coupling_2019} or radiatively\cite{young_picosecond_2012}. These processes are characterized by the recovery times $\tau_1$ and $\tau_2$. Both $\tau_1$ and $\tau_2$ recovery constants of $\Delta I^p_\mathrm{SHG} / I^p_\mathrm{SHG,0}$ are larger than those of $\Delta R/R_0$ because the dynamics of the spontaneous polarization results from the convolution of the faster dynamics of photoexcited carriers (seen by tr-refl) and the slower dynamics of atoms.

To interpret the SHG intensity drop and the reflectivity increase, it is useful to express the spontaneous polarization as\cite{yang_photostriction_2018, gattinoni_interface_2020}: $P_s(t) = 1/V \sum_i q_i(t) \Delta z_i(t)$, where $V$ is the volume of the unit cell, $q_i(t)$ is the Born effective charge and $\Delta z_i(t)$ is the out-of-plane displacement of atom $i$. The above-bandgap photoexcitation transfers electrons from the O 2p-derived valence band to the Ti 3d-derived conduction band of BTO. This charge transfer from O to Ti atoms reduces the corresponding charges $q_i$. We attribute changes in $R$ to changes in photoexcited carrier density $n_e$\cite{jin_structural_2012, sheu_ultrafast_2012, wang_coupling_2019}, thus contributing to $q_i$, while $P_s$ results from changes in both $q_i$ and $\Delta z_i$. After \SI{350}{\fs}, before atomic movements can occur, the increase in carrier density $n_e$ is responsible for the measured decrease in $P_s$ and increase in $R$, shown in Figures \ref{fig:Figure_3_shg_refl}e-f. This interpretation is corroborated by the increase of the maximum relative change of both $\Delta I^p_\mathrm{SHG}/I^p_\mathrm{SHG,0}$ and $\Delta R/R_0$ with pump fluence (\ref{si:fig:fluence_scan} and \ref{si:fig:time_trace_shg_refl_vs_fluence}).

\clearpage

\section*{Discussion}
\begin{figure}[h]
\centering
\includegraphics[width=1\textwidth]{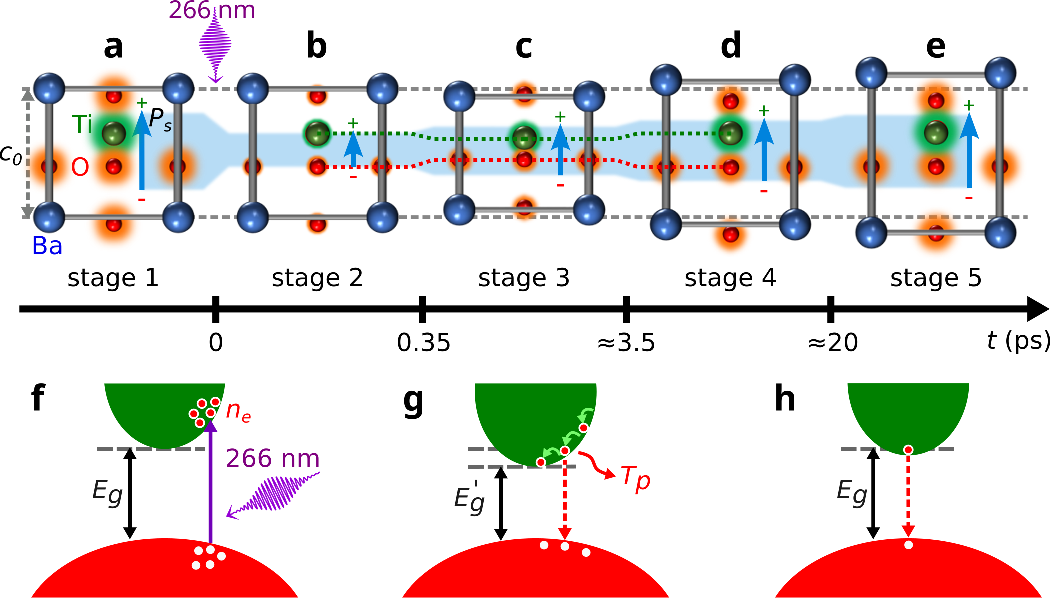}
\caption{\textbf{Schematic model of photoinduced lattice structure, polarization and electron dynamics.} BTO unit cell, projected along the [100] direction, at the following time delays $t$: $t < \SI{0}{\ps}$, i.e., ground state (a), $\SI{0}{\ps} < t < \SI{0.35}{\ps}$ (b), $\SI{0.35}{\ps} < t < \SI{3.5}{\ps}$ (c), $\SI{3.5}{\ps} < t < \SI{20}{\ps}$ (d), $\gtrapprox \SI{20}{\ps}$ (e). The distance between the horizontal dashed gray lines indicates the out-of-plane lattice parameter at equilibrium $c_0$ in panels (a)-(e). The size of the green and red shaded areas around Ti and O atoms, are proportional to the corresponding positive and negative Born effective charges, respectively. For clarity, the Born effective charge of Ba atoms is omitted. The length of the light blue arrow and the height of the blue shaded area are proportional to the magnitude of the spontaneous polarization $P_s$. Sketch of BTO valence and conduction band in stage 2 (f), in stages 3 and 4 (g), and in stage 5 (h). The purple vertical arrow indicates the photoexcitation of electrons from the valence to the conduction band upon absorption of a \SI{266}{\nm} laser photon. $E_g = \SI{3.4}{\eV}$ is the bandgap of the ground state, $E_g^{'}$ is the bandgap of the excited state, modified by the deformation potential, with $E_g^{'} - E_g \approx \SI{-3.2}{\milli\eV}$ (\ref{si:note:estimation_bandgap}). The relaxation of the hot electrons to the bottom of the conduction band (green arrows) and the recombination with holes of the valence band (red dashed vertical arrow) induce an increase of $T_p$. Panel (h) indicates the minor contribution of the deformation potential, resulting in a bandgap close to $E_g$, and the residual presence of electrons in the conduction band.}
\centering
\label{fig:Figure_4}
\end{figure}

We present here a unified picture in five stages of the lattice, polarization, and electron dynamics data presented above. Our observations are summarized in Figures \ref{fig:Figure_4}a-e, and the physical mechanisms involved are sketched in Figures \ref{fig:Figure_4}f-h. Before the arrival of the pump laser, the BTO is characterized by an out-of-plane polarization with given Born effective charges at each atomic site (stage 1, Figure \ref{fig:Figure_4}a). Upon absorption of the pump laser at $t = \SI{0}{\ps}$, electrons move from the occupied O-derived valence band to the unoccupied Ti-derived conduction band (Figure \ref{fig:Figure_4}f). Within $\approx \SI{350}{\fs}$ we observe the maximum increase in the photoexcited carrier density $n_e$, as indicated by the increase in $\Delta R / R_0$. This charge transfer reduces the Born effective charges $q_i$ at Ti and O atoms, thereby decreasing the spontaneous polarization $P_s$, as indicated by the decrease in $\Delta I_\mathrm{SHG}^p / I_\mathrm{SHG,0}^p$ (stage 2, Figure \ref{fig:Figure_4}b). A smaller $P_s$ is also consistent with a larger screening of long-range Coulomb interactions, which favor off-center atomic displacements and thus are responsible for the polar order. This effect is modeled by DFT calculations\cite{wang_ferroelectric_2012, paillard_photoinduced_2019}, showing that an increase in photoexcited carriers in the conduction band of BTO indeed tends to induce a phase transition from the ferroelectric to the paraelectric phase. Moreover, TD-DFT calculations\cite{lian_indirect_2019,chen_ferroelectric_2024} predict that such photoinduced change of BTO electronic structure weakens more significantly Ti-O$_\parallel$ bonds, parallel to $P_s$, as compared to Ti-O$_\perp$ bonds, perpendicular to $P_s$. We demonstrate this effect experimentally by measuring a larger change of the out-of-plane tensor elements ($\chi_{zxx}$, $\chi_{zzz}$) as compared to the in-plane tensor element ($\chi_{xxz}$). In stage 2, in contrast to the maximum change in carrier density $n_e$ and polarization $P_s$, the lattice remains unperturbed: this marks the onset of the decoupling between polarization and strain. At these early time delays, the bulk photovoltaic effect (BPVE) \cite{daranciang_ultrafast_2012} and the Schottky interface effect \cite{sarott_reversible_2024} could potentially play a role, but we explain in the following why they are not dominant in our experiments. 

First, in photoexcited ferroelectric materials it is common to observe the BPVE, i.e., the generation of photovoltage under light illumination \cite{fridkin_bulk_2001,dai_recent_2023,matsuo_bulk_2024}. The BPVE occurs under two conditions: the presence of a noncentrosymmetrical crystal and the excitation of nonthermalized electrons \cite{fridkin_bulk_2001}. In our system both conditions are satisfied shortly after $t = \SI{0}{\ps}$, i.e., before hot electron thermalization takes place (Figure \ref{fig:Figure_4}f). In BTO, if the light polarization is perpendicular to the spontaneous polarization $P_s$, the induced photovoltage is parallel to $P_s$ \cite{koch_bulk_1975,koch_anomalous_1976}. This enhances $P_s$ and induces, via the inverse piezoelectric effect, lattice expansion, similarly to what was observed in BiFeO$_3$ \cite{kundys_light-induced_2010}. However, in our study, although the polarization of the pump laser is perpendicular to $P_s$ (Figure \ref{fig:Figure_3_shg_refl}a), within $\approx \SI{350}{\fs}$, when electrons are non-thermal and BFVE could potentially play a role, we see a decrease in $P_s$ (Figure \ref{fig:Figure_3_shg_refl}e). Hence, we conclude that BPVE is not a dominant effect in our experiments. In fact, this observation can be explained by the wavelength dependence of the BPVE, which shows a maximum when the photon energy is close to $E_g$, as observed, e.g., in Refs. \cite{daranciang_ultrafast_2012,young_first_2012}. In contrast, in BTO at $\lambda = \SI{266}{\nm}$ the contribution of BVPE is negligible, as observed in experiments \cite{koch_bulk_1975,koch_anomalous_1976} and calculations \cite{dai_recent_2023}.

Second, ferroelectric thin films grown on a metal substrate form a Schottky interface, where the electron-hole pairs generated by light absorption are separated by the built-in voltage of the Schottky barrier and cause a power-independent enhancement of the polarization for $P_s$ pointing up (toward the surface) \cite{sarott_reversible_2024}. This effect can be excluded as responsible for our experimental observations because: (i) although our sample has $P_s$ pointing up (\ref{SI:fig:pfm}), we see a decrease in $P_s$, and (ii) the measured change in $P_s$ linearly depends on the incident fluence (\ref{si:fig:fluence_scan}). In fact, the Schottky interface effect is expected to play a significant role for films of thickness $\approx$ \SI{10}{\nm} or thinner \cite{sarott_reversible_2024}, and to be negligible in our case due to the three-fold greater film thickness.

In the $\SI{350}{\fs}-\SI{3.5}{\ps}$ delay range (stage 3, Figure \ref{fig:Figure_4}c), we observe polarization and strain following opposite trends. In fact, while the displacement between Ti and the center of the O octahedron as well as the strain decreases, $P_s$ starts to increase. This is attributed to the initial recombination of photoexcited carriers, suggested by the incipient recovery in $\Delta R / R$. At this stage, we also observe lattice compression caused by the negative parameter $\partial E_g/\partial p$ via the deformation potential. Our observation is in line with theory \cite{sanna_barium_2011, paillard_photoinduced_2019, chen_ferroelectric_2024} predicting a reduced bandgap $E_g$ due to the presence of photoinduced carrier density (Figure \ref{fig:Figure_4}g). Moreover, lattice contraction via the inverse piezoelectric effect is also consistent with an overall reduced $P_s$ caused by photoexcited carriers \cite{paillard_photostriction_2016}.

In the $3.5-7$ \SI{}{\ps} delay range (stage 4, Figure \ref{fig:Figure_4}d), the relative displacement between Ti and O atoms tends to restore the ground state with a larger $P_s$. At the same time, the relaxation of photocarriers to the top of the conduction band and their recombination result in a further increase of the Born effective charges, and an energy transfer to the phonon temperature (Figure \ref{fig:Figure_4}g), which, in turn, leads to lattice expansion. The latter effects persist up to $t \approx \SI{20}{\ps}$.

At $t \approx \SI{20}{\ps}$ (stage 5, Figure \ref{fig:Figure_4}e), together with a further relaxation of the electronic system toward equilibrium, we observe a saturation of the BTO average strain. This results in a metastable state with a slightly reduced $P_s$ and a significantly increased out-of-plane strain with respect to the ground state. The reduced $P_s$ is attributed to the presence of residual photoexcited carriers in the conduction band (Figure \ref{fig:Figure_4}h), while the increased out-of-plane strain is due to the thermoelastic contribution. We exclude the depolarization field screening as the main driving effect for the transient tensile strain seen in our experiments, because it would lead to the typical saturation of the strain with increasing pump fluence \cite{wen_efficient_2007}. For example, in PbTiO$_3$ \cite{daranciang_ultrafast_2012} the metastable lattice strain reaches saturation with \SI{266}{\nm} laser fluences of \SI{0.01}{\milli\joule\per\square\cm}. Conversely, in our study, we employ two orders of magnitude larger pump fluence and still the metastable strain scales approximately linearly with the pump fluence (Figure \ref{fig:Figure_1}d and \ref{SI:fig:eta_vs_delay_at_1_4mJ}). In addition, on the $\approx \SI{20}{\ps}$ timescale, the deformation potential plays a minor role (\ref{si:fig:TE_DP_strain}), thus its influence on the bandgap is expected to be marginal. Interestingly, although the out-of-plane lattice parameter is larger than in the ground state ($c_0$), the polarization $P_s$ is still smaller compared to equilibrium. This underscores the persistence of the decoupling between polarization and strain. Quantitatively, we estimate that in stage 5 the contribution of photoexcited electrons to $P_s$ has $\approx 10 \%$ larger magnitude than the structural contribution from the strain $\eta$ (\ref{si:note:estimation_electronic_contribution}).

In summary, we demonstrate that, with an above-bandgap laser excitation, an ultrafast decoupling of spontaneous polarization and strain can be achieved and measured. We assign this effect to the dominant contribution of photoexcited electrons in determining the spontaneous polarization when the system is out of equilibrium, and show that for an accurate description and fundamental understanding of a ferroelectric material in a photoexcited state, it is essential to combine multiple techniques that address the dynamical evolution of the various degrees of freedom, i.e., lattice, ferroelectric polarization, and electrons. The decoupling of polarization from the strain offers the opportunity to change paradigm from strain engineering \cite{li_insights_2024} to light-induced polarization engineering, thereby lifting the constraint of selecting among a limited number of substrates \cite{choi_enhancement_2004} or designing freestanding membranes \cite{pesquera_beyond_2020} to tune the spontaneous polarization. Moreover, by softening the Ti-O$_\parallel$ bonds, we bring BTO to an excited state, where it could be further modified by \SI{}{\tera\hertz} light to achieve stable and reversible polarization switching at lower fluences than otherwise needed when starting from the ground state \cite{mankowsky_ultrafast_2017}. 
Finally, the transient and reversible control of the spontaneous polarization, shown in this study, offers a pathway to control by light both electric and magnetic degrees of freedom in multiferroic materials.

\clearpage

\section*{Methods}
\label{sec:methods}

\textbf{Sample preparation and characterization}
\label{Methods:sample}

The epitaxial bilayers BTO/SRO were grown on a GSO substrate using pulsed laser deposition. The ceramic targets of SRO and BTO were \SI{8}{\cm} away from the substrate and ablated using a KrF excimer laser ($\lambda = \SI{248}{\nm}$, fluence \SI{5.4}{\joule\per\square\cm}, \SI{2}{\hertz} repetition rate). The deposition of SRO and BTO layers was conducted in O$_2$ atmosphere with pressure pO$_2$ = 100 mTorr and deposition temperatures of \SI{908}{\kelvin} and \SI{973}{\kelvin}, respectively. Sample cooling with the rate of \SI{3}{\kelvin\per\minute} was conducted in the environment of saturated O$_2$ (pO$_2$ = $10^4$ mTorr) to prevent the formation of oxygen vacancies. The thicknesses of BTO and SRO layers, $d_\mathrm{BTO} = \SI{34.5}{\nm}$ and $d_\mathrm{SRO} = \SI{47}{\nm}$ are extracted from a $\theta$-$2\theta$ scan of the as-grown sample around the (002) reflections (\ref{SI:fig:theta2theta}), while the GSO substrate is \SI{0.5}{\mm} thick. We determine the out-of-plane lattice parameters $c_\mathrm{BTO} = \SI{4.074}{\angstrom}$, $c_\mathrm{SRO} = \SI{3.934}{\angstrom}$, and $c_\mathrm{GSO} = \SI{3.964}{\angstrom}$ by means of the reciprocal space map shown in \ref{SI:fig:RSM}. The in-plane lattice parameter $a = \SI{3.970}{\angstrom}$, common to BTO, SRO and GSO, indicates that both BTO and SRO thin films are coherently strained to the substrate. Furthermore, the absence of satellite peaks in \ref{SI:fig:RSM} suggests the existence of a BTO monodomain. We determine that the spontaneous polarization of the as-grown sample $P_s$ points upward (toward the surface) by means of piezoresponse force microscopy (\ref{SI:fig:pfm}). The linear expansion coefficients of BTO above $T_c = \SI{400}{\celsius}$, of SRO and GSO in the temperature range $\SI{35}{\celsius} < T < \SI{700}{\celsius}$ are determined by $\theta$-$2\theta$ scans as a function of sample temperature (\ref{SI:fig:c_vs_temperature}).

\textbf{Time-resolved X-ray diffraction}
\label{Methods:tr-XRD}

Time-resolved X-ray diffraction experiments were performed at the Spectroscopy and Coherent Scattering instrument (SCS) of the European X-Ray Free-Electron Laser Facility (EuXFEL), using an optical laser (OL) as pump and the XFEL as probe. The XFEL pulse pattern consisted of pulse trains at a repetition rate of \SI{10}{\hertz}, with $35$ pulses per train at the intratrain repetition rate of \SI{113}{\kilo\hertz}. The full width at half maximum (FWHM) of the XFEL spectrum was \SI{11.7}{\eV}. To reduce the energy bandwidth, the XFEL beam was monochromatized using a variable line spacing grating with $50$ lines/mm in the first diffraction order and exit slits with a gap of \SI{100}{\micro\m}, providing an energy resolution of $\approx \SI{650}{\milli\eV}$. The nominal pulse duration of the XFEL pulses was $\approx \SI{25}{\fs}$, with pulse stretching at the monochromator grating of $\approx\SI{10}{\fs}$ (FWHM). The XFEL pulses, with initial energy of \SI{1.5}{\milli\joule} per pulse, were then attenuated by transmission through a gas attenuator (GATT), consisting of a volume containing $\mathrm{N_2}$ gas at a variable pressure. In order to prevent detector saturation, the transmission of the GATT was set to have $\approx \SI{15}{\nano\joule}$ per pulse at the sample. The XFEL pulse energy was measured by an X-ray gas monitor detector (XGM), located $\approx \SI{7}{\metre}$ upstream of the sample, just before the Kirkpatrick-Baez (KB) mirrors. The latter were used to focus the XFEL beam at the sample to a spot size of $w^\mathrm{XFEL}_x \times w^\mathrm{XFEL}_y = \SI{140}{\micro\m} \times \SI{100}{\micro\m}$ (determined by knife edge scans), where $w$ is defined as the distance from the beam axis where the intensity drops to $1/e^2$ of the value on the beam axis. The angle of incidence (defined from the sample surface) of XFEL and OL beams at the sample was \SI{86}{\degree} and \SI{85}{\degree}, respectively. The intensity of each XFEL pulse diffracted by the sample was measured by a Si avalanche photodiode (APD, model SAR3000G1X, Laser Components), converted to a voltage pulse and digitized. To prevent the pump laser intensity from reaching the APD, the latter was equipped with a filter made of \SI{400}{\nm} Ti, deposited on \SI{200}{\nm} polyimide. 

The pump laser had a central wavelength of \SI{266}{\nm}, the same pulse pattern as the XFEL with pulse duration of \SI{70}{\fs}, and beam size $w^\mathrm{OL}_x \times w^\mathrm{OL}_y = \SI{330}{\micro\m} \times \SI{240}{\micro\m}$, determined by knife edge scans. The transmittance profile of the \SI{266}{\nm} laser in our BTO/SRO/GSO sample is reported in \ref{si:note:transmittance_profile}. The incident pump laser fluence at the sample was \SI{2.7}{\mJ\per\square\cm}, and we verified that diffraction curves $I_\mathrm{XRD}(E_\nu)$ measured at negative delays $t$ coincide with those measured without the pump laser (\ref{si:fig:neg_delays_vs_laser_off}). This confirms the reversibility of the pump effect induced by UV laser light. The temporal overlap between XFEL and OL was determined as detailed in \ref{si:fig:t0}. The diffracted intensity of the XFEL pulses in a train was averaged and the corresponding time delay of the OL was corrected by the respective bunch arrival monitor (BAM) value, obtaining a time resolution $\Delta t \approx \SI{90}{\fs}$ (\ref{SI:note:BAM}). Data measured with incident pump laser fluence of \SI{1.4}{\mJ\per\square\cm} are reported in \ref{si:note:strain_model} and \ref{si:note:temp_strain_diffr_calculations}.

In general, two kinds of tr-XRD experiments were performed: photon energy scans at a fixed pump-probe delays $t$ (Figure \ref{fig:Figure_1}b and \ref{si:fig:diffraction_curve_drop}), and time delay scans at a fixed photon energy $E_\nu = \SI{1525}{\eV}$ (Figure \ref{fig:Figure_1}c and \ref{si:fig:t0}). The energy scans were carried out by a simultaneous movement of the monochromator grating and the undulators gap, such to have always the peak of the XFEL spectrum at the desired photon energy. From the energy scans at different $t$, the out-of-plane lattice parameters $c(t)$ and $c_0$ are calculated as the center-of-mass of the BTO diffraction intensity $I_\mathrm{XRD}(t)$ and refer to the average $c$ over $d_\mathrm{BTO}$ ($\overline{c}$). Specifically, the BTO average out-of-plane parameters $\overline{c}$ is calculated from the corresponding (001) Bragg peaks using the Bragg condition $\overline{c} = (\SI{12400}{\eV\angstrom})/(2\overline{E}_\nu\sin\theta)$, where $\overline{E}_\nu$ is the average of energy values, around the (001) BTO peak, weighted by $I_\mathrm{XRD}(E_\nu)$. Energy scans at different pump-probe delays $t$ over the photon energy range covering also GSO (001) and SRO (001) Bragg peaks are reported in \ref{si:fig:BTO_GSO_SRO_peaks_3mJ}.

\textbf{Time-resolved SHG and reflectivity}
\label{Methods:tr-SHG_tr-refl}

Time-resolved Second Harmonic Generation and time-resolved reflectivity experiments were performed at the SCS instrument of the EuXFEL using the same optical laser employed for tr-XRD experiments and the same pulse pattern. A sketch of the setup is shown in Figure \ref{fig:Figure_3_shg_refl}a. The \SI{800}{\nm} probe beam at frequency $\omega$ (red arrow) impinges on the sample at angle $\theta = \SI{50}{\degree}$ (defined from the normal to the surface) with polarization defined by the angle $\varphi$ and varied by rotating a half waveplate. The angle $\varphi = \SI{0}{\degree}$ [$\varphi = \SI{90}{\degree}$] refers to $p$ [$s$] polarized light. The \SI{266}{\nm} pump beam at frequency $3\omega$ (purple arrow) impinges on the sample at normal incidence with $p$ polarization. The \SI{800}{\nm} probe beam is then reflected by the sample and a dichroic mirror before reaching a Si photodiode. The latter is used to measure tr-refl of our sample. The \SI{400}{\nm} beam at frequency $2\omega$ (blue arrow) is the SHG signal generated in the BTO sample (\ref{SI:fig:shg_vs_800}). This SHG beam is transmitted through the dichroic mirror and Glan polarizer, which is set to select either the $p$ or $s$ component of the electric field, corresponding to the $p$-out or $s$-out configuration, respectively. Finally, the SHG beam is filtered by a \SI{400}{\nm} bandpass filter before reaching a photomultiplier (model H10721-210-Y004, Hamamatsu). The azimuthal rotation of the sample around the $z$ axis is defined by the angle $\gamma$. The independence of both polar plots of $I_\mathrm{SHG}^{p}(\varphi)$ and $I_\mathrm{SHG}^{s}(\varphi)$ from $\gamma$ confirms the out-of-plane nature of the spontaneous polarization of our BTO sample (\ref{SI:fig:polar_azimuth}). Both Si photodiode and photomultiplier measure respectively the reflectivity and SHG signal of each OL pulse in the train. The pulse duration of the \SI{800}{\nm} probe and \SI{266}{\nm} pump beams were approximately \SI{50}{\fs} and \SI{70}{fs}, providing a time resolution of $\approx \SI{90}{\fs}$. The beam sizes of \SI{800}{\nm} and \SI{266}{\nm} beams, determined by knife edge scans, were $w^{\SI{800}{\nm}}_x \times w^{\SI{800}{\nm}}_y = \SI{55}{\micro\m} \times \SI{46}{\micro\m}$ and $w^{\SI{266}{\nm}}_x \times w^{\SI{266}{\nm}}_y = \SI{165}{\micro\m} \times \SI{311}{\micro\m}$, respectively. The penetration depths of \SI{800}{\nm}, \SI{400}{\nm} and \SI{266}{\nm} laser beams in BTO are reported in \ref{si:note:penetration_depths}. While the fluence of the \SI{800}{\nm} probe beam was kept at \SI{1.3}{\mJ\per\square\cm}, the fluence of the \SI{266}{\nm} pump beam was set to \SI{2.7}{\mJ\per\square\cm} for the data displayed in Figure \ref{fig:Figure_3_shg_refl}. tr-SHG and tr-refl data with fluences between \SI{1.4}{\mJ\per\square\cm} and \SI{12.3}{\mJ\per\square\cm} are reported in \ref{si:fig:fluence_scan} and \ref{si:fig:time_trace_shg_refl_vs_fluence}.

In general, given the incoming electric field with components $E_j(\omega)$ and $E_k(\omega)$ at frequency $\omega$ along $j$ and $k$ directions, the electric dipole polarization induced in the material at frequency $2\omega$ along the direction $i$ is $P_i(2\omega) = \chi^{(2)}_{ijk}E_j(\omega)E_k(\omega)$, where $\chi^{(2)}_{ijk}$ is the second-order susceptibility, i.e., a third rank tensor reflecting the symmetry of the material. Each direction $i$, $j$, $k$ can be $x$, or $y$, or $z$ directions (Figure \ref{fig:Figure_3_shg_refl}a). By selecting $p$ or $s$ polarization of the SHG beam, we measure $I_\mathrm{SHG}^{p}(\varphi)$ or $I_\mathrm{SHG}^{s}(\varphi)$, which, for a BTO single crystal with 4mm point group symmetry, can be expressed as\cite{denev_probing_2011, sarott_reversible_2024}:
\begin{equation}\label{eq:I_shg_P}
I_\mathrm{SHG}^{p}(\varphi) \propto (\chi_{zxx}\sin{\theta}\cos{\varphi}^2  + (2\chi_{xxz}\cot{\theta}^2 + \chi_{zxx}\cot{\theta}^2 + \chi_{zzz}) \sin{\theta}^3\sin{\varphi}^2)^2,
\end{equation}
\begin{equation}\label{eq:I_shg_S}
I_\mathrm{SHG}^{s}(\varphi) \propto (2\chi_{xxz}\sin{\theta}\sin{\varphi}\cos{\varphi})^2.
\end{equation}
The resulting ratios of the tensor elements in Figure \ref{fig:Figure_3_shg_refl}b-c, $\chi_{zzz}^{(2)}/\chi_{zxx}^{(2)}=3.7$ and $\chi_{zzz}^{(2)}/\chi_{xxz}^{(2)}=6.5$, reflect a thin film under tensile out-of-plane stress\cite{zhao_stress-induced_2000}. The good quality of the fit curves in Figure \ref{fig:Figure_3_shg_refl}b-c confirms that our BTO thin film has 4mm point group symmetry. The minor discrepancies between data and fit model might be due to the fact that our BTO thin film is coherently strained to the substrate, and this leads to the appearance of additional minor nonzero tensor elements \cite{jeong_strain-induced_2000}.

\section*{Data availability}

Data recorded for the experiment at the European XFEL are available at doi:10.22003/XFEL.EU-DATA-003481-00. Source data are provided with this paper. Further experimental data as well as the simulation data that support the findings of this study are available from the corresponding author upon reasonable request.

\section*{Code availability}
Codes generated during the current study are available from the corresponding author on reasonable request.

\clearpage


\clearpage

\section*{Acknowledgements}
We acknowledge the European XFEL in Schenefeld, Germany, for provision of the XFEL beamtime at the SCS scientific instrument and would like to thank the staff for their assistance. D.P. acknowledges funding from 'la Caixa'  Foundation fellowship (ID 100010434) and the Spanish Ministry of Industry, Economy and Competitiveness (MINECO), grant no PID2019-109931GB-I00. The ICN2 is funded by the CERCA programme/Generalitat de Catalunya and by the Severo Ochoa Centres of Excellence Programme, funded by the Spanish Research Agency (AEI, CEX2021-001214-S). T.C.A. acknowledges funding from the Heisenberg Resonant Inelastic X-ray Scattering (hRIXS) Consortium. We thank D. Hickin for support with the automation of tr-SHG measurements. We thank A. Reich and J.T. Delitz for the design of mechanical components used in tr-XRD experiments. We are grateful to M. Altarelli for careful reading of the manuscript.

\section*{Author contributions}
G.Merc., with input from L.P.H., G.C., I.A.V., J.Z. and T.L.L., conceived the experiments. G.C. and I.S., with input from G.Merc., designed the sample. J.M.C.R manufactured the sample and D.P. characterized it with $\theta$-$2\theta$ scan, RSM and PFM. L.P.H, D.P., G.N.H., R.C., L.M., M.T., S.G., T.C.A., G.Merz., S.P., J.Sc., Z.Y., I.A.V. and G.Merc. performed tr-XRD experiments. C.C. developed an online analysis tool to visualize tr-XRD data in real time. G.Merc. and L.P.H. performed tr-SHG and tr-refl experiments. J.Sa. performed $\theta$-$2\theta$ measurements at different sample temperatures. L.P.H. and G.Merc. analyzed all the data. L.P.H. modeled tr-XRD and tr-SHG data. G.Merc., A.S., K.R. and I.A.V. supervised the project. G.Merc. wrote the manuscript, drafted by L.P.H., and with input from all authors. All authors provided critical feedback and helped shape the research, analysis and manuscript.

\section*{Ethics declarations}
The authors declare no competing interests.

\clearpage

\begin{figure}[h]
\centering
\includegraphics[width=1\textwidth]{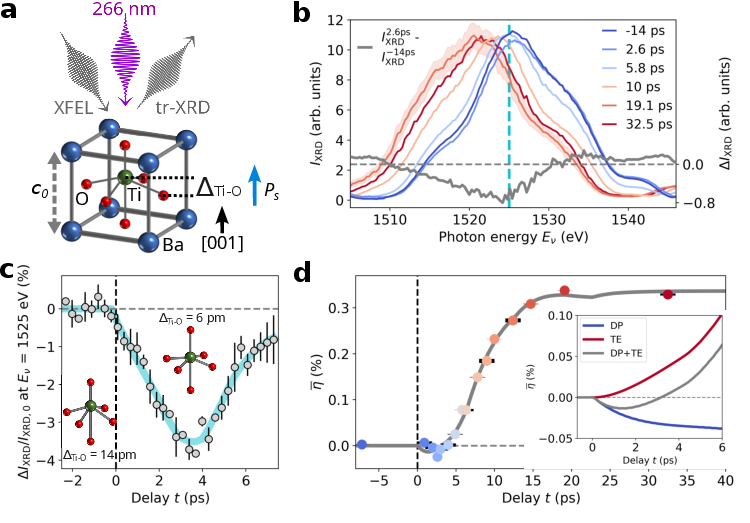}
\caption{\textbf{Photoinduced structural dynamics and time evolution of strain.} (a) Sketch of the BaTiO$_3$ unit cell, the out-of-plane lattice parameter in the ground state $c_0$, the relative displacement $\Delta_\mathrm{Ti-O}$ of the Ti atom from the center of the O octahedron, the spontaneous ferroelectric polarization $P_s$, the pump laser at $\SI{266}{\nm}$, the incident and diffracted XFEL beams. (b) Selected experimental $I_\mathrm{XRD}(E_\nu)$ curves at different time delays $t$, and the difference $\Delta I_\mathrm{XRD} = I_\mathrm{XRD}^{\SI{2.6}{\ps}} - I_\mathrm{XRD}^{\SI{-14}{\ps}}$ (gray line). The vertical dashed cyan line marks $E_\nu = \SI{1525}{\eV}$. The shaded red area indicates the standard deviation value $\mathrm{SD} \approx 0.1 \times I_\mathrm{XRD}^{\SI{19.1}{\ps}}$, with similar values also for the diffraction curves at different delays $t$. For clarity, only the error bar of $I_\mathrm{XRD}^{\SI{19.1}{\ps}}$ is shown. (c) Delay dependence of $\Delta I_\mathrm{XRD} / I_\mathrm{XRD,0} = [I_\mathrm{XRD}(t) - I_\mathrm{XRD,0}]/I_\mathrm{XRD,0}$, i.e., the relative change of $I_\mathrm{XRD}(t)$ at the photon energy $E_\nu=\SI{1525}{\eV}$ with respect to the equilibrium value at negative delays $I_\mathrm{XRD,0}$. The solid cyan line is a guide to the eye. The error bars indicate the standard deviation of experimental data in a bin size of \SI{300}{\fs}. Inset: sketch of Ti and O octahedron at negative delays ($\Delta_\mathrm{Ti-O} = \SI{14}{\pm}$), and at \SI{3.5}{\ps} ($\Delta_\mathrm{Ti-O} = \SI{6}{\pm}$). (d) Average BTO out-of-plane strain $\overline{\eta}(t)$ as a function of pump-probe delay $t$. The error bars follow from the determination of $c$ from $I_\mathrm{XRD}(E_\nu) \pm \mathrm{SD}$. The solid gray line is a fit to the data (\ref{si:note:strain_model}). Inset: Modeled contributions to the average strain $\overline{\eta}(t)$ due to deformation potential (DP) and thermoelastic (TE) stress (\ref{si:fig:TE_DP_strain}).}
\centering
\label{fig:Figure_1}
\end{figure}

\clearpage

\begin{figure}[h]
\centering
\includegraphics[width=0.7\textwidth]{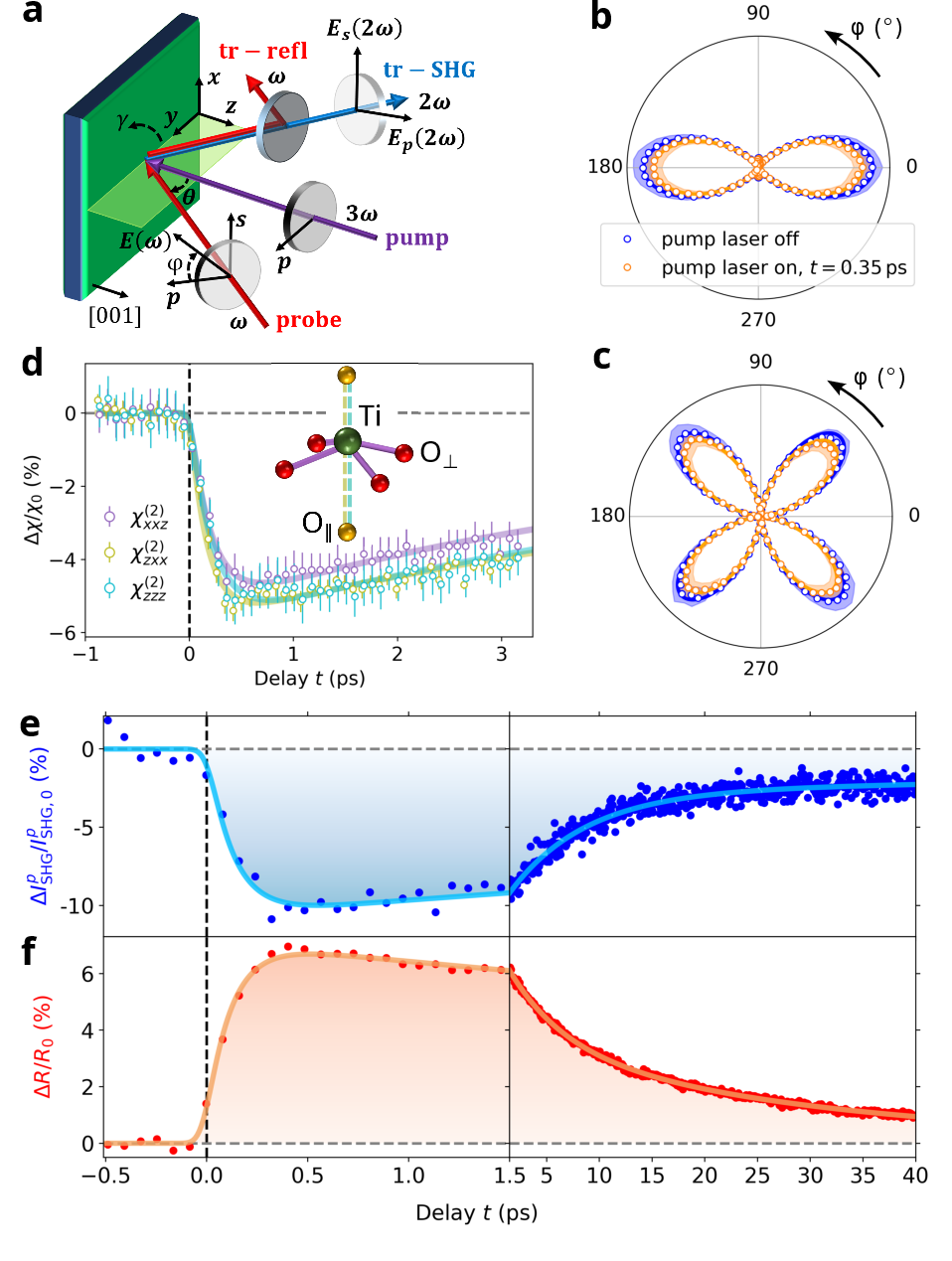}
\caption{\textbf{Photoinduced ferroelectric polarization and electron dynamics.} (a) Sketch of the tr-SHG and tr-refl setup. Polar plots of $I_\mathrm{SHG}^{p}(\varphi)$ (b) and $I_\mathrm{SHG}^{s}(\varphi)$ (c) measured without pump laser (blue points) and with pump laser at the delay $t=\SI{0.35}{\ps}$ (orange points). The green solid lines are fit curves to the data resulting from equations \eqref{eq:I_shg_P} and \eqref{eq:I_shg_S}. The shaded orange and blue areas refer to the standard deviation of the data and amount to $\approx 13\%$. (d) Relative change $\Delta \chi/\chi_0$ of the tensor elements $\chi_{xxz}^{(2)}$, $\chi_{zxx}^{(2)}$, and $\chi_{zzz}^{(2)}$ as a function of delay $t$, and respective fit curves (\ref{SI:note:fit_function}). $\Delta \chi/\chi_0= (\chi^{(2)}_{ijk} - \chi^{(2)}_{ijk,0})/\chi^{(2)}_{ijk,0}$, where $\chi^{(2)}_{ijk,0}$ refers to the tensor element $\chi^{(2)}_{ijk}$ at $t \leq \SI{0}{\ps}$. The error bars refer to the standard deviation resulting from the fit of the tensor elements. Inset: sketch of Ti atom and O octahedron with the indication of $\mathrm{O_\parallel}$ (yellow spheres) and $\mathrm{O_\perp}$ (red spheres), and the softening of Ti-$\mathrm{O_\parallel}$ bonds (dashed olive and cyan lines) with respect to the  Ti-$\mathrm{O_\perp}$ bonds (solid purple lines). Relative change of SHG $\Delta I^p_\mathrm{SHG}/I^p_\mathrm{SHG,0}$ (e) and reflectivity $\Delta R/R_0$ (f) as a function of the delay $t$. $\Delta I_\mathrm{SHG}^p/I^p_\mathrm{SHG,0} = [I_\mathrm{SHG}^p(t) - I_\mathrm{SHG,0}^p]/I_\mathrm{SHG,0}^p$, where $I_\mathrm{SHG,0}^p$ refers to the SHG intensity at $t \leq \SI{0}{\ps}$. $\Delta R/R_0 = [R(t) - R_0]/R_0$, where $R_0$ refers to the reflectivity at $t \leq \SI{0}{\ps}$. The solid lines are fit curves to the data with fit parameters $\tau_0$, $\tau_1$, and $\tau_2$ reported in the text. The error bar of $\Delta I^p_\mathrm{SHG}/I^p_\mathrm{SHG,0}$ and $\Delta R/R_0$ data points are $\approx 8\%$ and $2\%$, respectively. Since SHG is a nonlinear process, it is more significantly affected by fluctuations of the $\SI{800}{\nm}$ probe laser intensity of $\approx 2\%$. However, the standard deviation of $\Delta I^p_\mathrm{SHG} / I^p_\mathrm{SHG,0}$ and $\Delta R/R_0$ at $t<\SI{0}{\ps}$ are $0.6\%$ and $0.06\%$, respectively.}
\centering
\label{fig:Figure_3_shg_refl}
\end{figure}

\clearpage

\renewcommand{\figurename}{}
\renewcommand{\thefigure}{Figure S\arabic{figure}}
\setcounter{figure}{0}

\renewcommand \theequation{S\arabic{equation}}
\setcounter{equation}{0}

\renewcommand \thesection{Supplementary Note \arabic{section}}

\renewcommand{\tablename}{}
\renewcommand \thetable{Table S\arabic{table}}

\clearpage

\pagebreak
\hspace{0pt}
\vfill
\LARGE
  \textbf{Supplementary Information for: \\ Ultrafast decoupling of polarization and strain in ferroelectric $\mathrm{BaTiO_3}$}
\vfill
\hspace{0pt}
\pagebreak

\normalsize




\clearpage

\section{Sample characterization}

\begin{figure*}[h]
\centering
\includegraphics[width=0.8\textwidth]{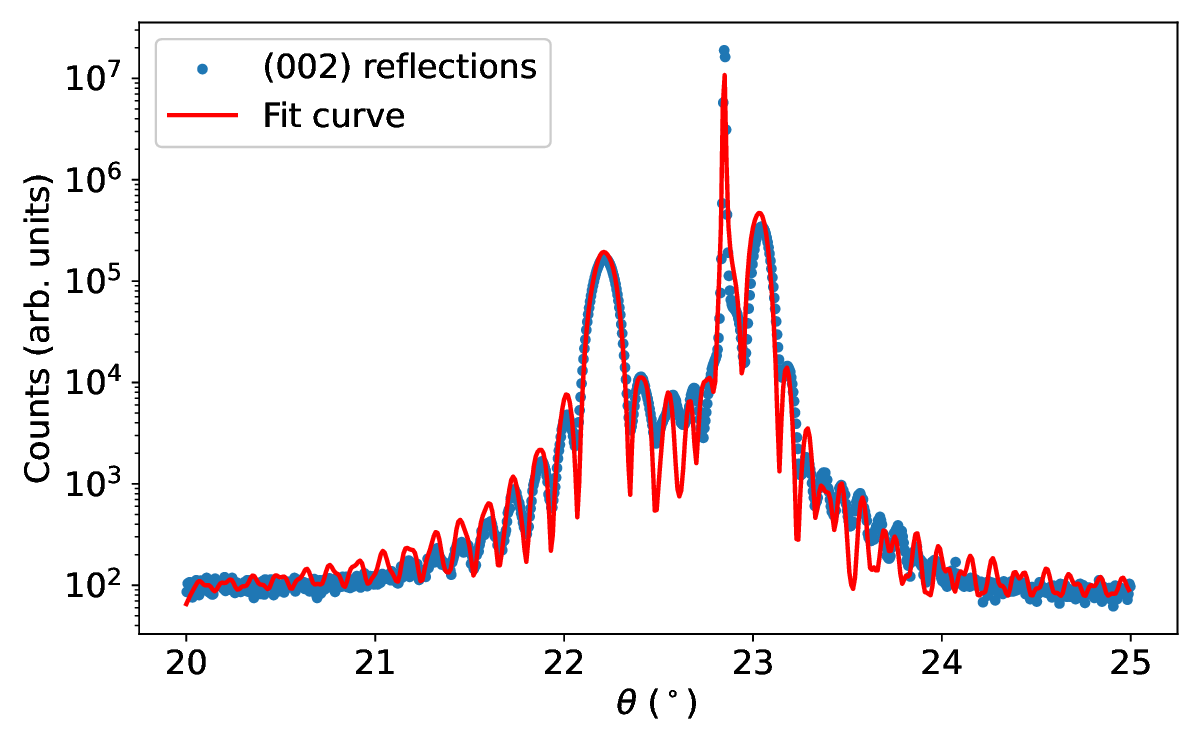}
\caption[$\theta$-$2\theta$ scan]{\textbf{$\theta$-$2\theta$ scan.} $\theta$-$2\theta$ scan of the as-grown sample around the (002) reflections of BTO and SRO thin films, and the GSO substrate. The measurements are performed using a PANalytical X'Pert Pro diffractometer, and the experimental data are modeled by the dynamical theory of diffraction with BTO and SRO thicknesses as fit parameters, providing $d_\mathrm{BTO} = \SI{34.5}{\nm}$ and $d_\mathrm{SRO} = \SI{47}{\nm}$, respectively.}
\centering
\label{SI:fig:theta2theta}
\end{figure*}

\clearpage

\begin{figure*}[h]
\centering
\includegraphics[width=0.6\textwidth]{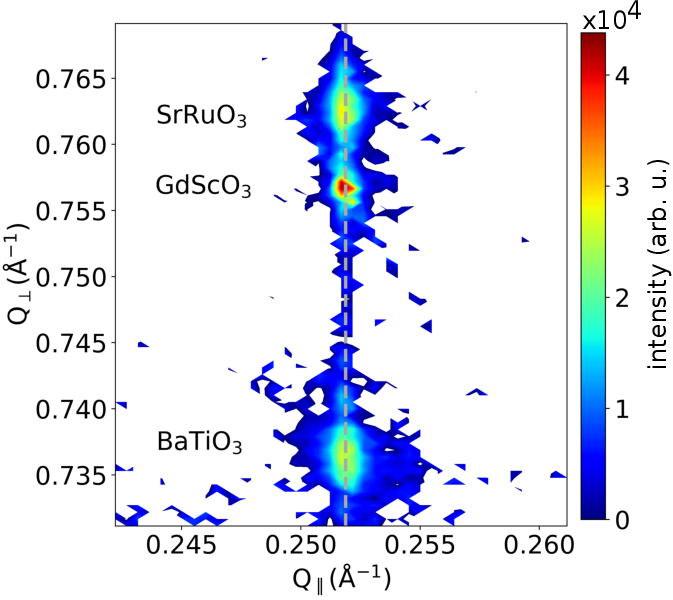}
\caption[Reciprocal space map]{\textbf{Reciprocal space map.} X-ray reciprocal space map (RSM) of the  sample around the ($\overline{1}03$) substrate Bragg peak, measured by a PANalytical X'Pert Pro diffractometer. The vertical gray dashed line indicates the reciprocal lattice parameter $Q_\parallel$ shared by the GSO substrate, BTO, and SRO thin films. The reciprocal lattice parameters $Q_\parallel$ and $Q_\perp$ of the intensity peaks are related to the real space in-plane $a$ and out-of-plane $c$ lattice parameters by the following relations \cite{Mario05}: $a = Q_\parallel^{-1} = -\lambda/(2Q_\mathrm{x})$ and $c = (Q_\perp/3)^{-1} = (3\lambda)/(2Q_\mathrm{z})$. Here, $\lambda = \SI{1.5406}{\angstrom}$ is the wavelength of the incident Cu k$\alpha$ radiation, $Q_\mathrm{x} = \sin(\theta) \sin(\theta - \omega)$, and $Q_\mathrm{z} = \sin(\theta) \cos(\theta - \omega)$, where $2\theta$ is the angle between incident and outgoing wavevector, and $\omega$ is the angle between incident wavevector and sample surface. Fitting the RSM intensity distribution with a pseudo-Voigt function \cite{Young82} provides $Q_\parallel$ and $Q_\perp$ of each diffraction peak, and the following lattice parameters: $a = \SI{3.970}{\angstrom}$, $c_\mathrm{BTO} = \SI{4.074}{\angstrom}$, $c_\mathrm{SRO} = \SI{3.934}{\angstrom}$, and $c_\mathrm{GSO} = \SI{3.964}{\angstrom}$.}
\centering
\label{SI:fig:RSM}
\end{figure*}

\clearpage

\begin{figure*}[h]
\centering
\includegraphics[width=0.4\textwidth]{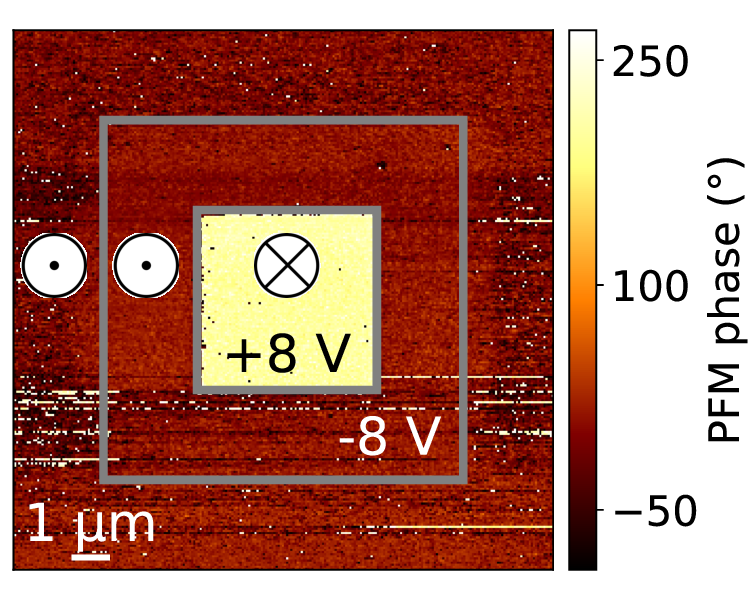}
\caption[Piezoresponse force microscopy]{\textbf{Piezoresponse force microscopy.} Piezoresponse force microscopy (PFM) phase image of the BTO/SRO/GSO sample. Piezoresponse force microscopy in Dual AC Resonance Tracking (DART) mode \cite{rodriguez_dual-frequency_2007} is used to probe the polarization of as-grown samples and to prove that ferroelectric polarization can be switched by the application of positive or negative voltage between the PFM tip and the SRO electrode. The application of tip voltage \SI{+8}{\volt} [\SI{-8}{\volt}], to the regions marked by the gray boxes, forces a downward [upward] polarization $P_s$ pointing away from the surface ($\otimes$) [toward the surface ($\odot$)], with the corresponding PFM phase $180\degree$ [$0\degree$]. The PFM phase in the region where no voltage is applied (beyond the outer gray box) reveals that our as-grown sample has upward polarization $P_s$ with no indication of multiple domains.}
\centering
\label{SI:fig:pfm}
\end{figure*}

\clearpage

\begin{figure*}[h]
\centering
\includegraphics[width=1\textwidth]
{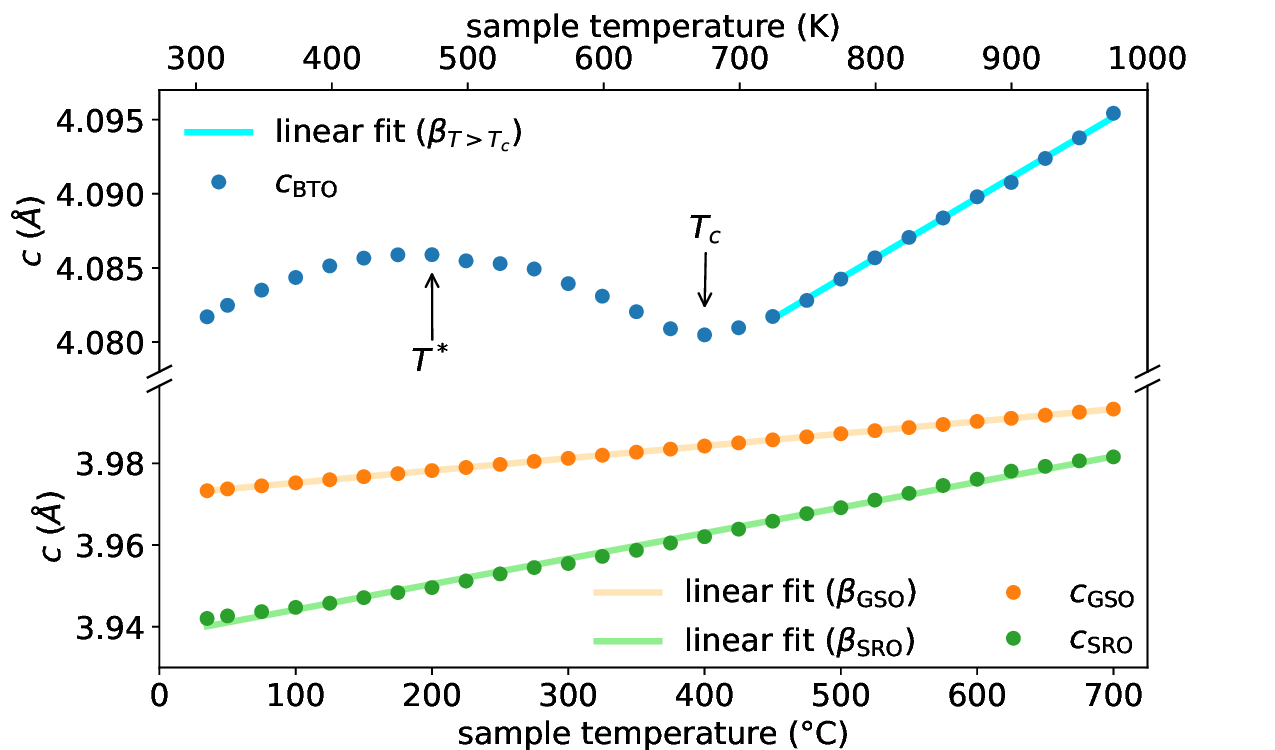}
\caption[$c$ parameters as a function of sample temperature]{\textbf{$c$ parameters as a function of sample temperature.} Out-of-plane lattice constant $c$ of BTO (blue points), SRO (green points), and GSO (orange points) as a function of the sample temperature. The $c$ parameters are extracted from $\theta$-$2\theta$ scans (similar to \ref{SI:fig:theta2theta}) measured at different sample temperatures. The arrows indicate $T^* = \SI{200}{\celsius}$ and $T_c = \SI{400}{\celsius}$. The blue line is a linear fit of $c_\mathrm{BTO}$ data that provides $\beta_{T>T_c} = \SI{1.33e-5}{\per\kelvin}$. The orange and green lines are linear fits of $c_\mathrm{GSO}$ and $c_\mathrm{SRO}$ data, which provide $\beta_\mathrm{GSO} = \SI{7.6e-6}{\per\kelvin}$ and $\beta_\mathrm{SRO} = \SI{1.58e-5}{\per\kelvin}$, respectively.}
\centering
\label{SI:fig:c_vs_temperature}
\end{figure*}

\clearpage

\section{Fit function of tr-XRD, tr-SHG and tr-refl delay scans}\label{SI:note:fit_function}

The function used to fit the delay scans presented in this work is:
\begin{equation}\label{si:eq:fit_function}
    F(t) = h(t) \ast g(t).
\end{equation}
Equation \eqref{si:eq:fit_function} is the convolution of the sum of three exponential decays $h(t) = \sum_{i=1}^{3}A_i\exp\big(-(t-t_0)/\tau_i\big)$, with the Gaussian function $g(t) = \sqrt{2}/(\sigma\sqrt{\pi})\exp\big(-(t-t_0)^2/(2\sigma^2)\big)$, where $\sigma = \Delta t / (2\sqrt{2\ln{2}})$ and $\Delta t$ is the time resolution (FWHM) of our experiment (\ref{SI:note:BAM}). The resulting fit function can be written as:
\begin{equation}\label{si:eq:fit_function_explicit}
    F(t) = \sum_{i=1}^{3}A_i\exp\left(\frac{\sigma^2}{2\tau_i^2}-\frac{t-t_0}{\tau_i}\right)\left(1-\erf\left(\frac{\sigma^2-\tau_i(t-t_0)}{\sqrt{2}\sigma \tau_i}\right)\right),
\end{equation}
with the time constants $t_0$, $\tau_1$, $\tau_2$, and $\tau_3$ and the amplitudes $A_1$, $A_2$, and $A_3$ as fit parameters.

\clearpage

\section{XRD data}

\begin{figure}[h]
\centering
\includegraphics[width=0.8\textwidth]{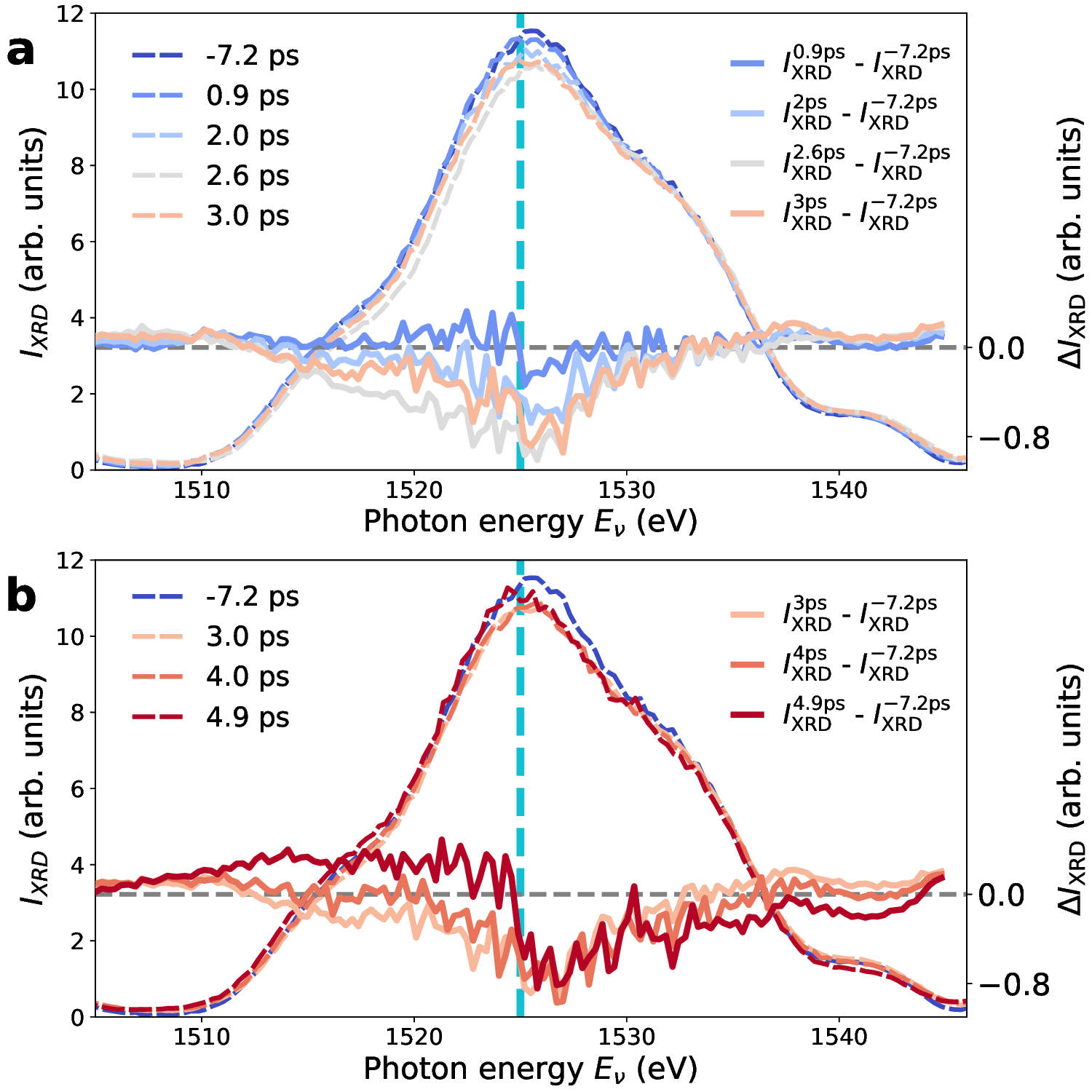}
\caption{\textbf{Diffraction intensity drop.} Experimental $I_\mathrm{XRD}(E_\nu)$ curves at different pump-probe delays $t$, and the respective differences $\Delta I_\mathrm{XRD} = I_\mathrm{XRD}^t - I_\mathrm{XRD}^{\SI{-7.2}{\ps}}$, where $t = \SI{0.9}{\ps}, \SI{2}{\ps}, \SI{2.6}{\ps}, \SI{3}{\ps}, \SI{4}{\ps}, \SI{4.9}{\ps}$. The vertical dashed cyan line marks $E_\nu = \SI{1525}{\eV}$. These data highlight that up to $\approx 3-4 \SI{}{\ps}$ there is mostly a drop in diffraction intensity near the peak center, accompanied by a minor shift of the spectral weight to higher photon energies ($I_\mathrm{XRD}^{\SI{2.6}{\ps}}$), indicating lattice compression. After \SI{4}{\ps} the diffraction intensity near the peak center increases and the peak shifts to the lower photon energies ($I_\mathrm{XRD}^{\SI{4.9}{\ps}}$), indicating lattice expansion.}
\centering
\label{si:fig:diffraction_curve_drop}
\end{figure}

\clearpage

\begin{figure}[h]
\centering
\includegraphics[width=1\textwidth]
{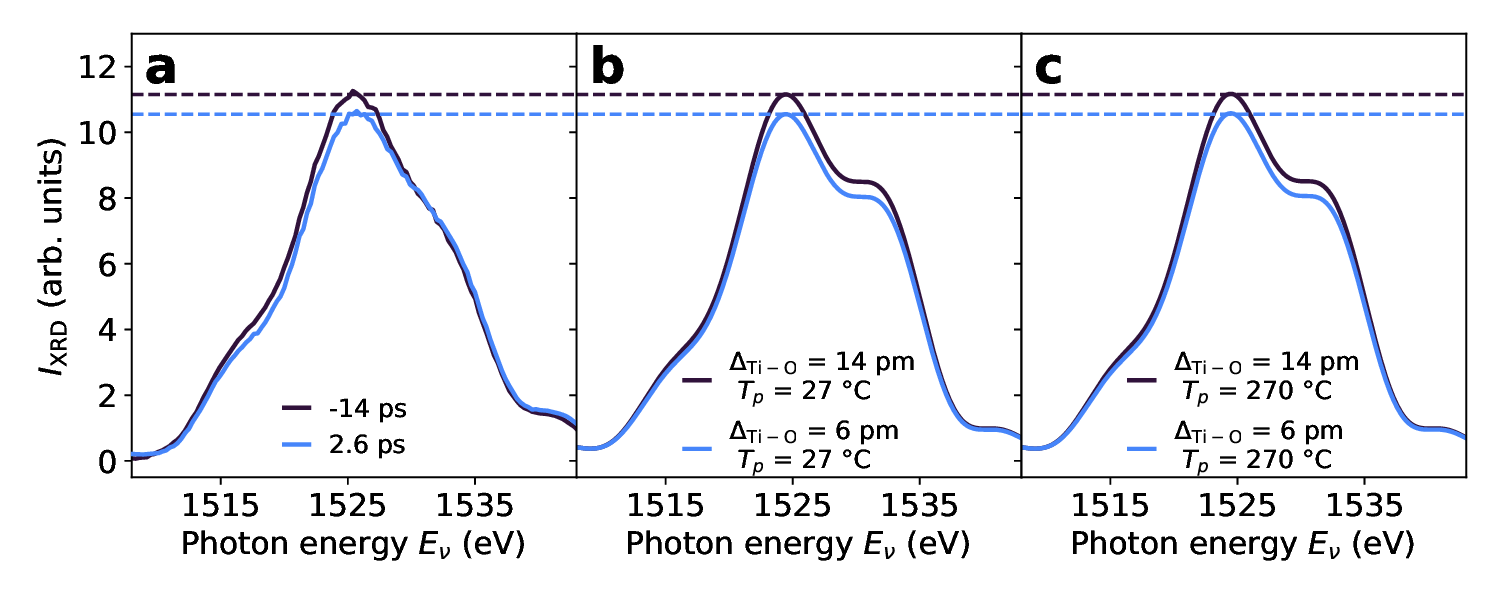}
\caption[Contribution of Debye-Waller factor]{\textbf{Contribution of Debye-Waller factor.} (a) BTO (001) diffraction peak measured at $t=\SI{-14}{\ps}$ and $t=\SI{2.6}{\ps}$ (same as in Figure 1b). Diffraction intensity simulations that mimic the drop of peak diffraction intensity with $\Delta_\mathrm{Ti-O} = \SI{14}{\pm}$ and $\Delta_\mathrm{Ti-O} = \SI{6}{\pm}$ at $T_p = \SI{27}{\celsius}$ (b) and $T_p = \SI{270}{\celsius}$ (c). The minimal difference between the diffraction curves in panels (b) and (c) shows that the effect of the average lattice temperature $T_p$ on the diffraction intensity, via the Debye-Waller effect, is negligible in the $T_p$ range relevant to this work ($T_p$ $\leq$ \SI{270}{\celsius}, \ref{si:fig:Temperature_profile}b). Moreover, the non-monotonic decrease and increase in peak diffraction intensity in the region $\SI{0}{\ps}<t<\SI{7}{\ps}$ is not consistent with the monotonic increase of $T_p$ upon absorption of the pump laser (\ref{si:fig:Temperature_profile}b).}
\centering
\label{si:fig:DW_effect}
\end{figure}

\clearpage

\begin{figure*}[h]
\centering
\includegraphics[width=1\textwidth]
{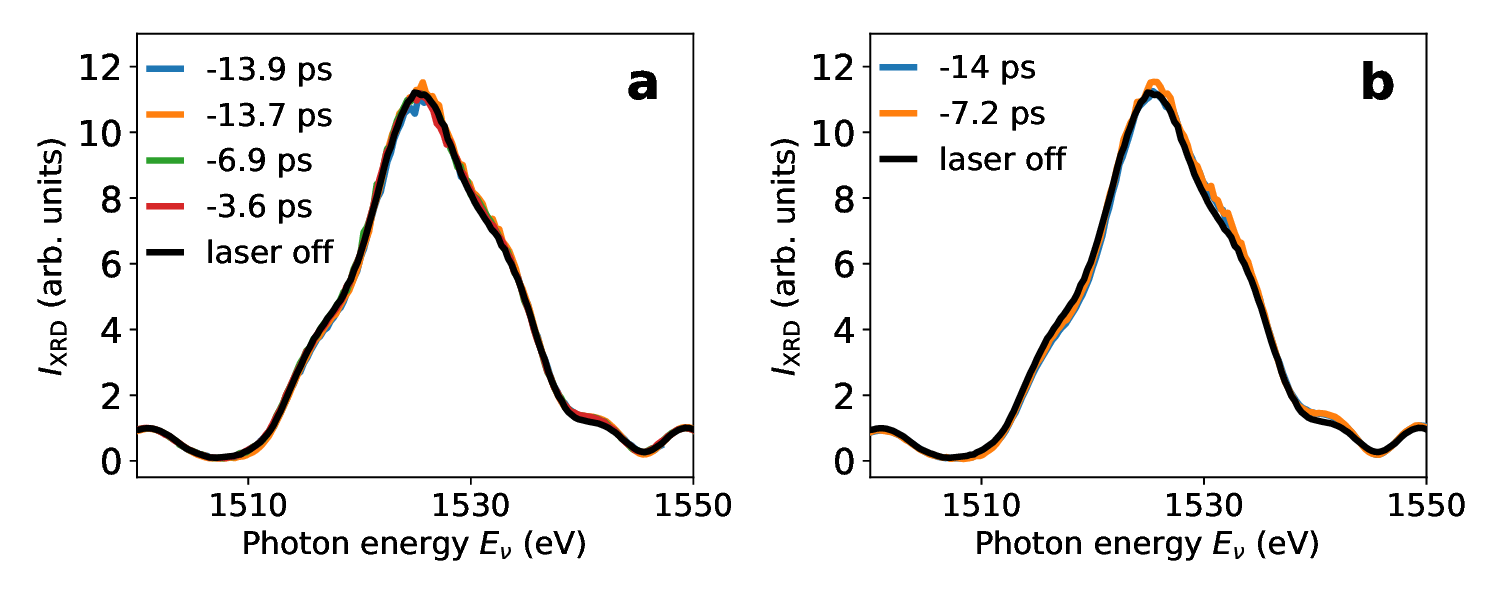}
\caption[Comparison of $I_\mathrm{XRD}(E_\nu)$ at negative delay and laser off]{\textbf{Comparison of $I_\mathrm{XRD}(E_\nu)$ at negative delay and laser off.} Comparison of laser off X-ray diffraction data $I_\mathrm{XRD}(E_\nu)$ with data measured at negative delays, at incident fluence $F_\mathrm{in} = \SI{1.4}{\mJ\per\square\cm}$ (a) and $F_\mathrm{in} = \SI{2.7}{\mJ\per\square\cm}$ (b). The difference among different diffraction curves is negligible.}
\centering
\label{si:fig:neg_delays_vs_laser_off}
\end{figure*}

\clearpage

\begin{figure*}[h]
\centering
\includegraphics[width=0.6\textwidth]
{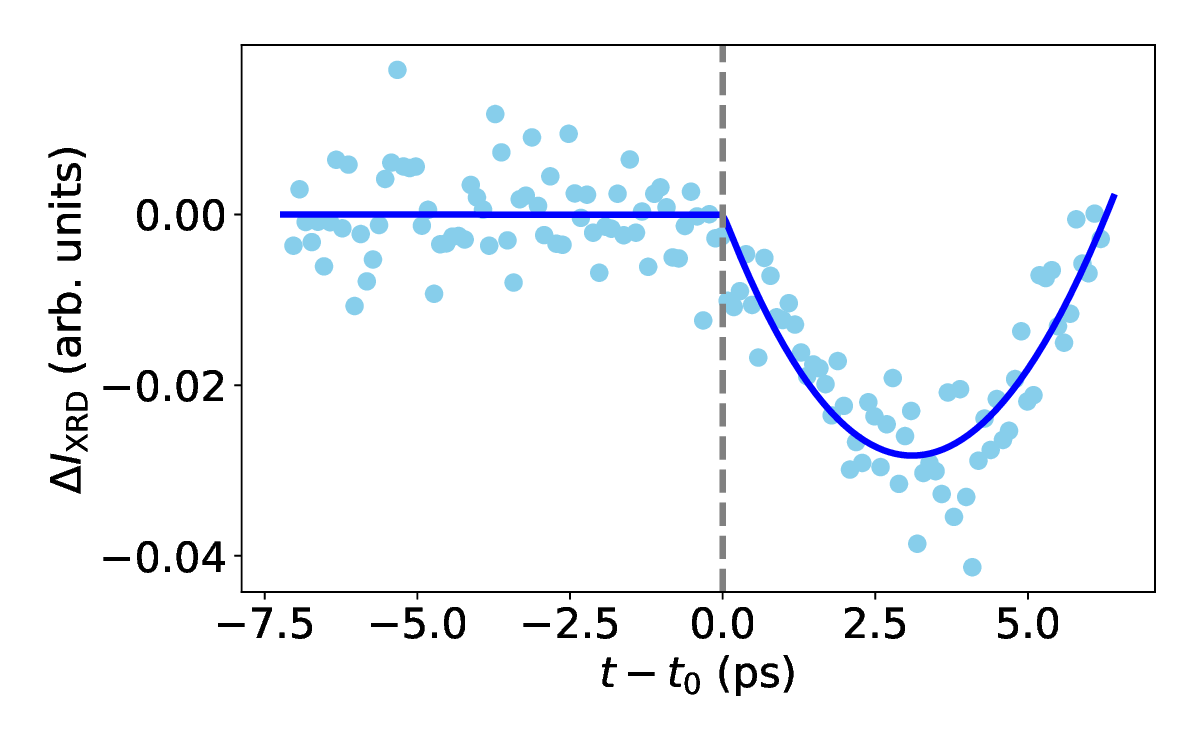}
\caption[Determination of $t_0$ in tr-XRD experiments]{\textbf{Determination of $t_0$ in tr-XRD experiments.} The time delay $t_0$ corresponding to the temporal overlap of XFEL and OL during the tr-XRD measurements was determined by fitting high statistics $\Delta I_\mathrm{XRD}(t-t_0)$ data at $E_\nu = \SI{1525}{\eV}$ (shown here) using the fit function \eqref{si:eq:fit_function_explicit} (\ref{SI:note:fit_function}).}
\centering
\label{si:fig:t0}
\end{figure*}

\clearpage

\begin{figure*}[h]
\centering
\includegraphics[width=1\textwidth]
{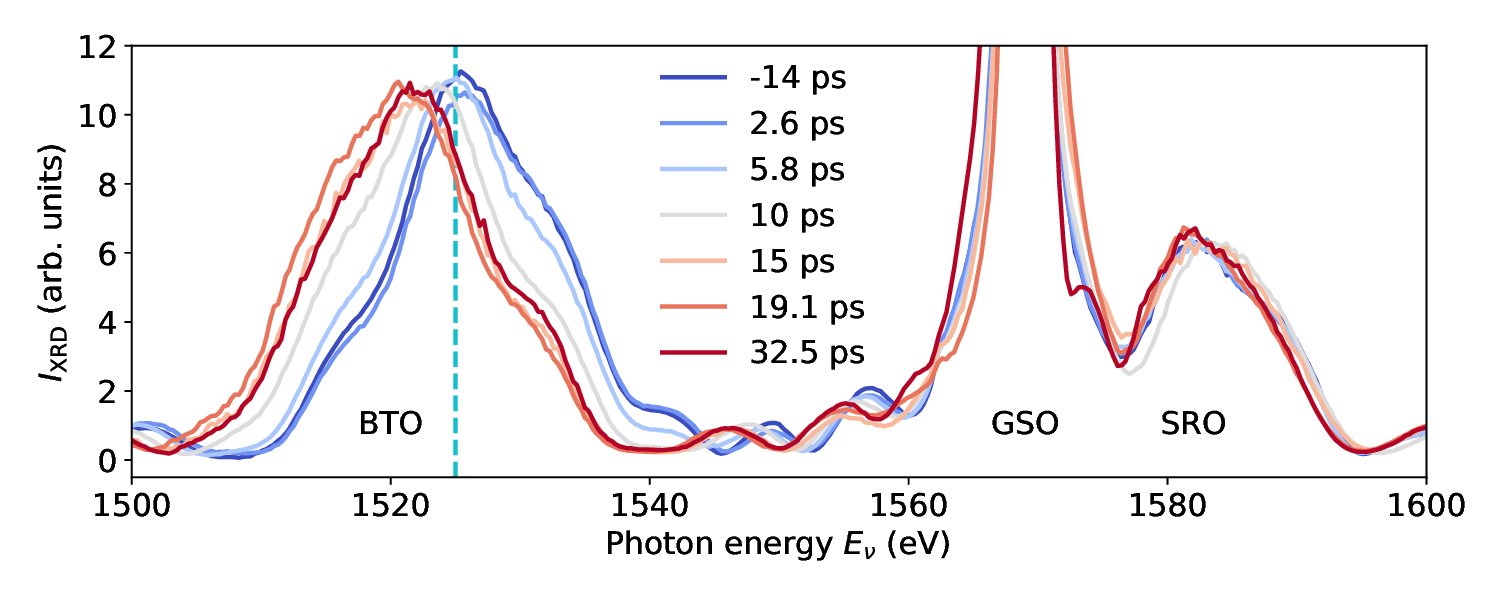}
\caption[BTO, SRO and GSO (001) diffraction peaks]{\textbf{BTO, SRO and GSO (001) diffraction peaks.} Time-resolved (001) Bragg peaks of the BTO/SRO/GSO sample at different time delays $t$. In contrast to the relatively large energy shifts of the BTO Bragg peak, the SRO and GSO (001) Bragg peaks show much weaker energy shifts, corresponding to changes of the average strain smaller than $0.05\%$ and $0.02 \%$, respectively.}
\centering
\label{si:fig:BTO_GSO_SRO_peaks_3mJ}
\end{figure*}

\clearpage

\section{Bunch arrival time monitor and time resolution of tr-XRD experiments}
\label{SI:note:BAM}

The bunch arrival time monitor (BAM) at the European XFEL tracks the arrival time of electron bunches in each pulse train, providing invaluable information to improve the time resolution of pump-probe experiments \cite{lohl_electron_2010, czwalinna_beam_2021}. The BAM measures the electron bunch arrival time with respect to the master clock that is also used to synchronize the pump laser. We employ the most downstream BAM, located at the end of the accelerator tunnel (\SI{1932}{\metre} from the laser gun) and about \SI{1.5}{\kilo\metre} upstream of the interaction point at the sample position. In previous time-resolved experiments at the SCS Instrument, it has been observed that the time drifts measured by the BAM are amplified by $17 \% - 45 \%$ at the interaction point \cite{carley_scs_2022}. In practice, this translates in a further correction of up to few tens of femtoseconds. In our experiments, the exact amplification coefficient is not known, thus we assume a direct proportionality between BAM and $t_0$ at the experiment, where $t_0$ indicates the temporal overlap between XFEL and optical laser. This assumption is justified by the dynamics of tr-XRD experiments taking place on the picosecond timescale (Figure \ref{fig:Figure_1}). 

\begin{figure*}[h]
\centering
\includegraphics[width=0.8\textwidth]
{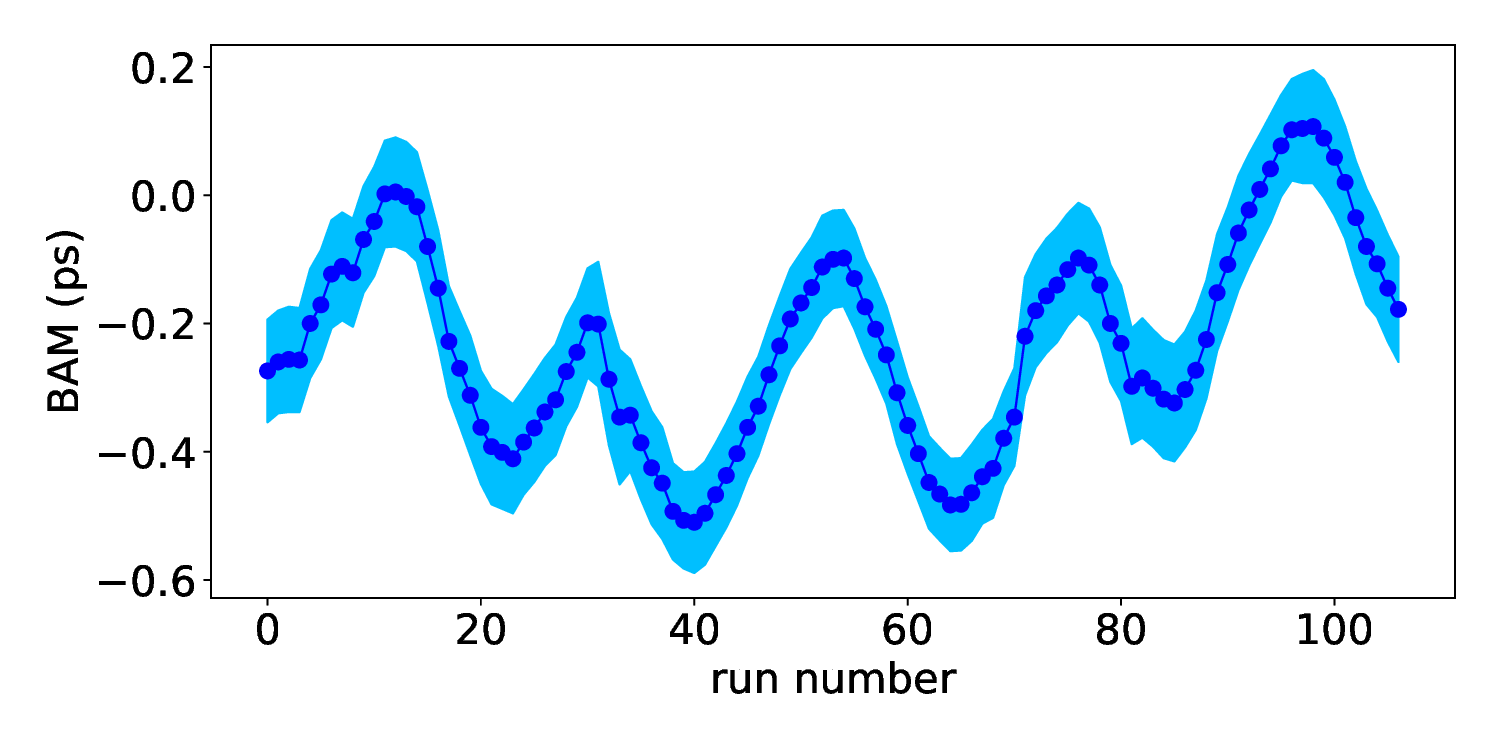}
\caption[Beam arrival monitor]{\textbf{Beam arrival monitor.} Mean (points) and standard deviation (shaded area) of the BAM values of consecutive runs (not evenly spaced in time) recorded within \SI{60}{\hour}.}
\centering
\label{SI:fig:BAM}
\end{figure*}

We measure a standard deviation of the BAM within a pulse train of \SI{10}{\fs}, and a standard deviation of the BAM from train to train of \SI{84}{\fs}. Most importantly, slow drifts of the arrival time of FEL pulses over the course of hours are present and shown in \ref{SI:fig:BAM} for each recorded run. This indicates that without BAM correction, the time resolution of our experiments might be limited up to $\approx \SI{600}{\fs}$, if we average non-consecutive runs. To improve our time resolution, the time delays of our data, indicating the relative time between FEL and optical laser pulses, are corrected for each pulse train according to the average BAM value of the respective pulse train. Thus, the resulting expected time resolution is $\Delta t = \sqrt{\tau_\mathrm{FEL}^2 + \sigma_{\tau_\mathrm{FEL}}^2 + \tau_\mathrm{OL}^2 + \sigma_{\tau_\mathrm{OL}}^2 } \approx \SI{90}{\fs}$. This assumes an FEL pulse duration of $\tau_\mathrm{FEL} \approx \SI{35}{\fs}$, including the nominal \SI{25}{\fs} pulse duration \cite{schneidmiller_photon_2010} and the pulse stretching at the monochromator of $\approx\SI{10}{\fs}$ (FWHM) \cite{gerasimova_soft_2022}, a pulse-to-pulse jitter in the train of $\sigma_{\tau_\mathrm{FEL}} \approx \SI{20}{\fs}$, an optical laser pulse duration of $\tau_\mathrm{OL} \approx \SI{70}{\fs}$ and temporal jitter of $\sigma_{\tau_\mathrm{OL}} \approx \SI{30}{\fs}$.

\clearpage

\section{SHG and reflectivity data}

\begin{figure*}[h]
\centering
\includegraphics[width=0.8\textwidth]{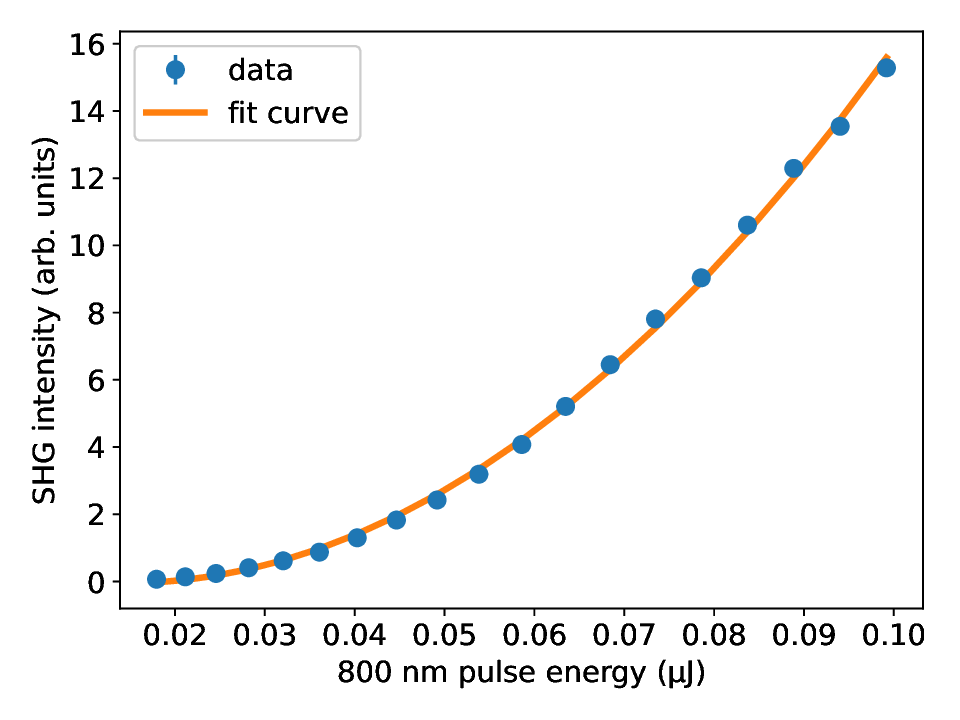}
\caption[SHG intensity as a function of probe pulse energy]{\textbf{SHG intensity.} SHG intensity measured as a function of the \SI{800}{\nm} probe pulse energy. The fit curve to the data is a polynomial of second order, thus it confirms the nonlinear nature of the signal measured by the photomultiplier.}
\centering
\label{SI:fig:shg_vs_800}
\end{figure*}

\clearpage

\begin{figure*}[h]
\centering
\includegraphics[width=1\textwidth]
{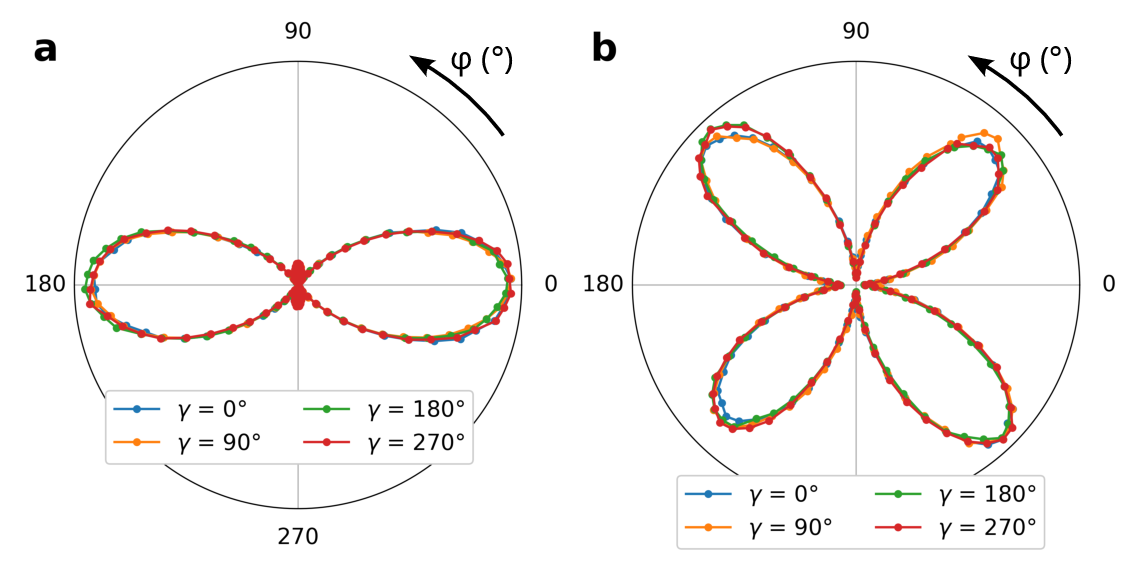}
\caption[SHG plots at different azimuthal angles]{\textbf{SHG plots at different azimuthal angles.} Polar plots in $p$-out (a) and $s$-out (b) configuration measured at the four azimuthal angles $\gamma = \SI{0}{\degree}, \SI{90}{\degree}, \SI{180}{\degree}, \SI{270}{\degree}$. Regardless of the angle $\gamma$, the polar plots measured in the same polarization configuration overlap. This confirms the out-of-plane nature of the spontaneous polarization of our BTO sample. The presence of an in-plane component would manifest as different $\varphi$-dependence for different angles $\gamma$ \cite{zhang_characterization_2018}.}
\centering
\label{SI:fig:polar_azimuth}
\end{figure*}

\clearpage

\begin{figure*}[h]
\centering
\includegraphics[width=0.7\textwidth]
{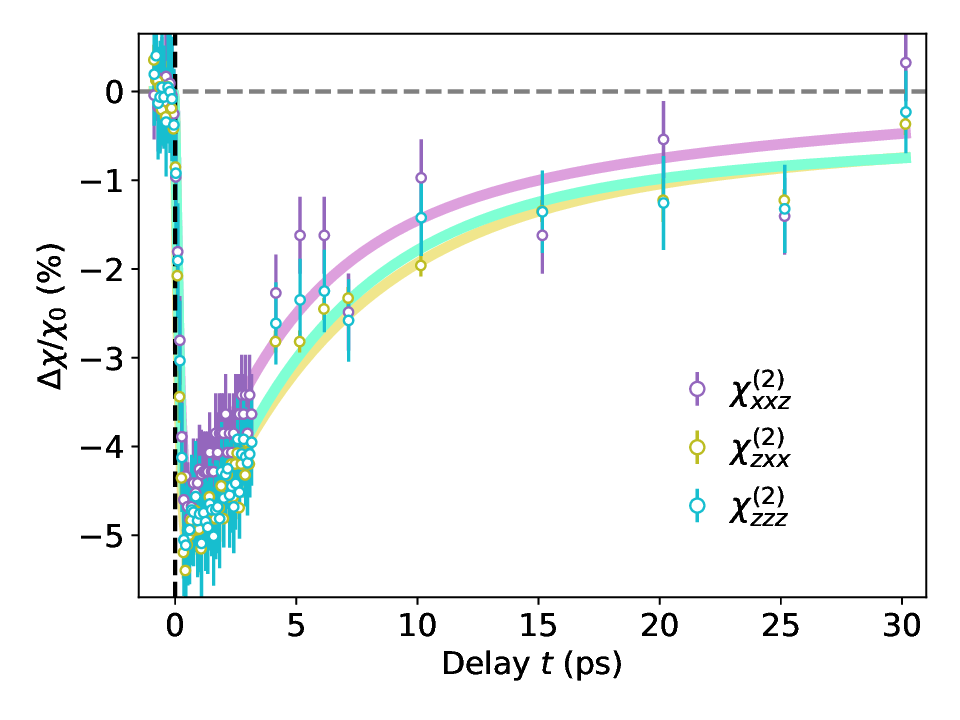}
\caption[Delay dependence of the tensor elements $\chi_{xxz}^{(2)}$, $\chi_{zxx}^{(2)}$, and $\chi_{zzz}^{(2)}$]{\textbf{Delay dependence of the tensor elements $\chi_{xxz}^{(2)}$, $\chi_{zxx}^{(2)}$, and $\chi_{zzz}^{(2)}$.} Relative change $\Delta \chi/\chi_0$ of the tensor elements $\chi_{xxz}^{(2)}$, $\chi_{zxx}^{(2)}$, and $\chi_{zzz}^{(2)}$ as a function of delay $t$, and corresponding fit curves given by equation \eqref{si:eq:fit_function_explicit} (\ref{SI:note:fit_function}). The fit results are: $\tau_0^{\chi_{xxz}}=\SI{200\pm50}{\fs}$, $\tau_1^{\chi_{xxz}}=\SI{4.2\pm2.2}{\ps}$, $\tau_2^{\chi_{xxz}}=\SI{23.5\pm28}{\ps}$, $\tau_0^{\chi_{zxx}}=\SI{160\pm30}{\fs}$, $\tau_1^{\chi_{zxx}}=\SI{6.1\pm2.4}{\ps}$, $\tau_2^{\chi_{zxx}}=\SI{49\pm119}{\ps}$,
$\tau_0^{\chi_{zzz}}=\SI{220\pm40}{\fs}$, $\tau_1^{\chi_{zzz}}=\SI{5.6\pm2.2}{\ps}$, $\tau_2^{\chi_{zzz}}=\SI{64\pm219}{\ps}$. The error bars refer to the standard deviation resulting from the fit of the tensor elements. The relatively small difference between $\chi_{zxx}^{(2)}$ (or $\chi_{zzz}^{(2)}$) and $\chi_{xxz}^{(2)}$ is due to the unfavorable experimental geometry with the polarization of the pump laser nearly perpendicular to $P_s$ (Figure \ref{fig:Figure_3_shg_refl}a), and it is expected to increase with the component of the pump polarization along $P_s$ \cite{chen_ferroelectric_2024}. }
\centering
\label{si:fig:fit_chi_long_delay_scans}
\end{figure*}

\clearpage

\begin{figure}[h]
\centering
\includegraphics[width=0.8\textwidth]
{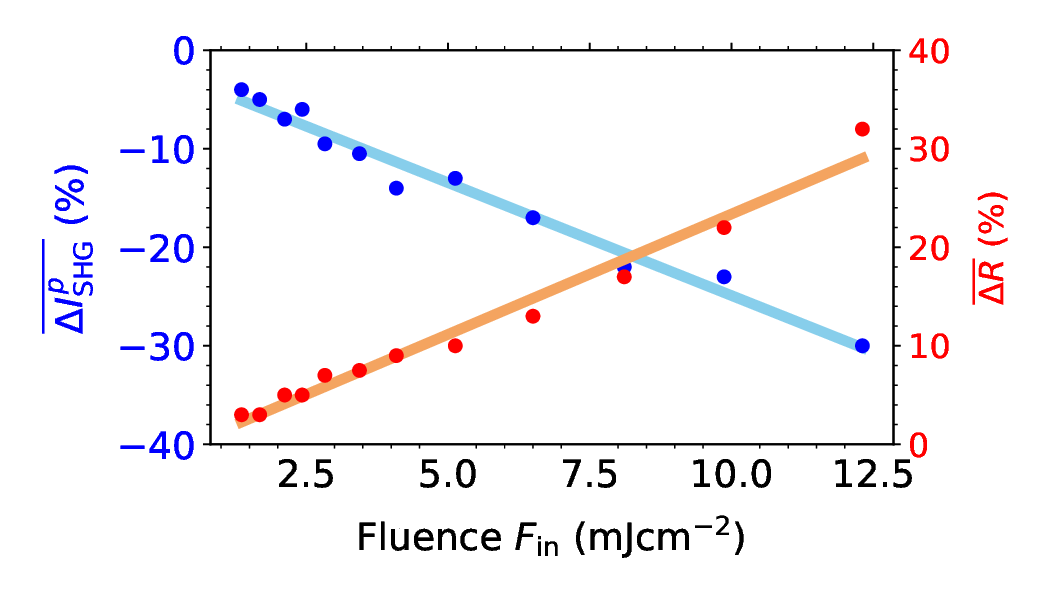}
\caption[SHG and $R$ maximum relative change]{\textbf{SHG and $R$ maximum relative change.} Maximum relative change of SHG ($\overline{\Delta I^p_\mathrm{SHG}}$, blue points) and reflectivity ($\overline{\Delta R}$, red points) as a function of the incident pump fluence $F_\mathrm{in}$. The parameters $\overline{\Delta I^p_\mathrm{SHG}}$ and $\overline{\Delta R}$ are the average over the delay range $\SI{0.32}{\ps}<t<\SI{0.73}{\ps}$ of $\Delta I^p_\mathrm{SHG} / I^p_\mathrm{SHG}$ and $\Delta R/R$, respectively (\ref{si:fig:time_trace_shg_refl_vs_fluence}). Blue and red lines are linear fits to the data. The error bars of each data point, equal to the standard deviation of $\Delta I^p_\mathrm{SHG}/I^p_\mathrm{SHG,0}$ and $\Delta R/R_0$ in the delay range $\SI{0.32}{\ps}<t<\SI{0.73}{\ps}$, are not visible as they are approximately $1\%$.}
\centering
\label{si:fig:fluence_scan}
\end{figure}

\clearpage

\begin{figure}[h]
\centering
\includegraphics[width=1\textwidth]
{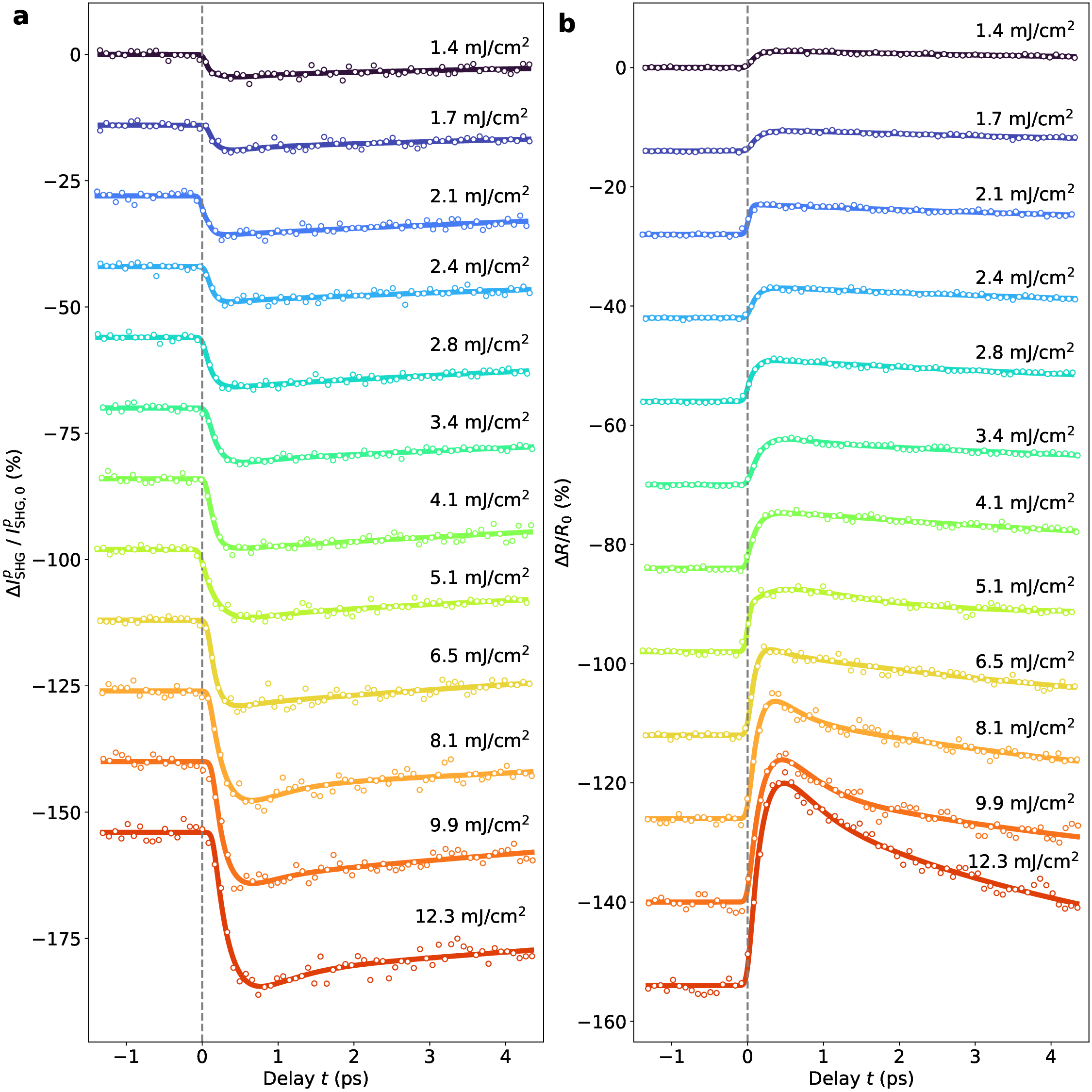}
\caption[SHG and $R$ time traces at different pump fluences]{\textbf{SHG and $R$ time traces at different pump fluences.} $\Delta I^p_\mathrm{SHG}/I^p_\mathrm{SHG,0}$ (a) and $\Delta R/R_0$ (b) as a function of pump-probe delay $t$, measured at different incident pump laser fluences $F_\mathrm{in}$ marked on the figure. $\Delta I^p_\mathrm{SHG} / I^p_\mathrm{SHG,0}$ data are measured at $\varphi = 0\degree$. The maximum relative change of SHG ($\overline{\Delta I^p_\mathrm{SHG}}$) and reflectivity ($\overline{\Delta R}$), shown in \ref{si:fig:fluence_scan} are derived from the data reported in this figure. Each time trace has an offset of $-14\%$ from the one above.}
\centering
\label{si:fig:time_trace_shg_refl_vs_fluence}
\end{figure}

\clearpage

\section{Sample properties}\label{si:note:sample_properties}

\subsection{Transmittance profile of \SI{266}{\nm} beam in BTO/SRO/GSO}\label{si:note:transmittance_profile}

\begin{figure*}[h]
\centering
\includegraphics[width=0.7\textwidth]
{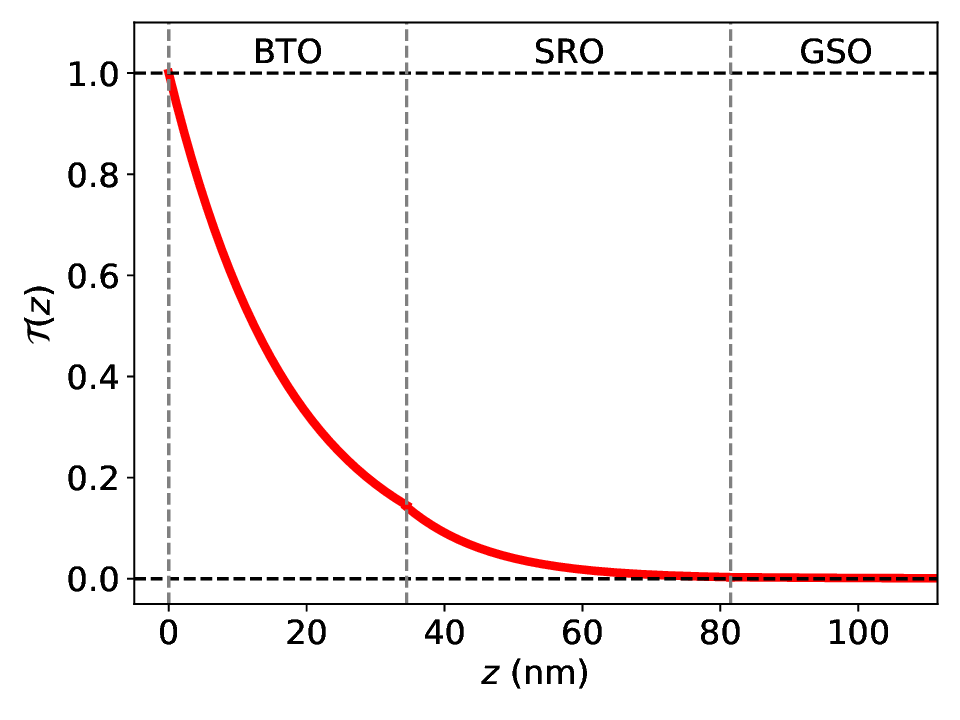}
\caption[Transmittance profile]{\textbf{Transmittance profile.} Transmittance profile $\mathcal{T}(z)$ of the \SI{266}{\nm} pump laser in our BTO/SRO/GSO sample employing pentration depths $\delta$ and reflectivities $R_p$ reported in \ref{SI:tab:constants}. The dashed gray lines mark the BTO surface, the BTO/SRO and the SRO/GSO interfaces.}
\centering
\label{SI:fig:transmittance_profile}
\end{figure*}

The transmittance of the \SI{266}{\nm} beam in our BTO/SRO/GSO sample, displayed in \ref{SI:fig:transmittance_profile}, is calculated as:
\begin{equation}
\begin{split}
   & \mathcal{T}_\mathrm{BTO}(z) = \exp(-z/(\delta_\mathrm{BTO}\cos{\theta_t^\mathrm{BTO}})), \mathrm{for}\,\, 0<z<d_\mathrm{BTO} \\  
   & \mathcal{T}_\mathrm{SRO}(z) = \mathcal{T}_\mathrm{BTO}(d_\mathrm{BTO}) (1-R_p^\mathrm{SRO}) \exp(-z/(\delta_\mathrm{SRO}\cos{\theta_t^\mathrm{SRO}})), \mathrm{for}\,\, d_\mathrm{BTO}<z<d_\mathrm{BTO}+d_\mathrm{SRO} \\
   & \mathcal{T}_\mathrm{GSO}(z) = \mathcal{T}_\mathrm{SRO}(d_\mathrm{BTO}+d_\mathrm{SRO}) (1-R_p^\mathrm{GSO}) \exp(-z/(\delta_\mathrm{GSO}\cos{\theta_t^\mathrm{GSO}})), \mathrm{for}\,\, z>d_\mathrm{BTO}+d_\mathrm{SRO}.
\end{split}   
\end{equation}

The penetration depths $\delta_\mathrm{BTO}$, $\delta_\mathrm{SRO}$ and $\delta_\mathrm{GSO}$ are calculated as detailed in \ref{si:note:penetration_depths}. The reflections at the air-BTO, BTO-SRO, and SRO-GSO interfaces (\ref{SI:tab:constants}) are determined by using Snell's law $n_1 \sin{\theta_i} = n_2 \sin{\theta_t}$ and the Fresnel equation for $p$-polarized light $R_p = |(n_1\cos{\theta_t} - n_2\cos{\theta_i})/(n_1\cos{\theta_t} + n_2\cos{\theta_i})|^2$. Here, $n_1$ and $n_2$ are the real part of $n^*$ at the two sides of the interface, while $\theta_i$ and $\theta_t$ are the angles of the incident and the transmitted beam with respect to the surface normal, with $\theta_i^\mathrm{BTO} = 4\degree$. The index of refraction of SRO and GSO are calculated from Ref. \cite{thompson_enhanced_2016}, as detailed in \ref{si:note:penetration_depths}, while the index of refraction of the BTO thin film is determined from ellipsometry data at \SI{266}{\nm} in Ref. \cite{chernova_strain-controlled_2015}.

Given the parameters above (summarized in \ref{SI:tab:constants}), the absorbed fluences in the BTO and SRO thin films are calculated as:
\begin{equation}
   F_\mathrm{abs}^\mathrm{BTO} = F_\mathrm{in}(1-R_p^\mathrm{BTO})(1-\exp(-d_\mathrm{BTO}/(\delta_\mathrm{BTO}\cos{\theta_t^\mathrm{BTO}}))) = 0.7 F_\mathrm{in}
\end{equation}

\begin{equation}
F_\mathrm{abs}^\mathrm{SRO} = F_\mathrm{in}(1-R_p^\mathrm{BTO})\exp[-d_\mathrm{BTO}/(\delta_\mathrm{BTO}\cos{\theta_t^\mathrm{BTO}})](1-R_p^\mathrm{SRO})[1-\exp(-d_\mathrm{SRO}/(\delta_\mathrm{SRO}\cos{\theta_t^\mathrm{SRO}}))] = 0.12 F_\mathrm{in},
\end{equation}
and are reported in \ref{si:tab:absorbed_fluences}.

\begin{table*}[h]
\caption[Absorbed fluences in BTO and SRO thin films]{\textbf{Absorbed fluences in BTO and SRO thin films.} Absorbed fluences in BTO and SRO thin films at two different incident fluences $F_\mathrm{in}$.}
\begin{tabular}{ccccc}
\hline
 $F_\mathrm{in}$ (\SI{}{\milli\joule\per\square\cm}) & $F_\mathrm{abs}^\mathrm{BTO}$ (\SI{}{\milli\joule\per\square\cm}) & $F_\mathrm{abs}^\mathrm{SRO}$ (\SI{}{\milli\joule\per\square\cm}) \\ 
\hline
  1.4 & $0.98$ & $0.17$ \\
\hline
  2.7 & $1.88$ & $0.33$ \\
\hline
\end{tabular}\label{si:tab:absorbed_fluences}
\end{table*}

\subsection{Penetration depths}\label{si:note:penetration_depths}

The absorption coefficient $\alpha$ of the \SI{266}{\nm} pump laser in our BTO thin film is determined based on the study of $\alpha$ as a function of strain in Ref. \cite{chernova_strain-controlled_2015}. For a compressive strain of $-0.55\%$, the penetration depth is $\delta_\mathrm{BTO} = 1/\alpha_\mathrm{BTO} = \SI{17.9}{\nm}$. The penetration depths in the SRO thin film $\delta_\mathrm{SRO} = \SI{21.5}{\nm}$ and in the GSO substrate $\delta_\mathrm{GSO} = \SI{26.1}{\nm}$ are determined from the respective dielectric constants $\epsilon_1$ and $\epsilon_2$, reported in Ref. \cite{thompson_enhanced_2016}. Specifically, $\delta = 1/\alpha = \lambda/(4\pi k)$, where $\lambda = \SI{266}{\nm}$ is the wavelength of the pump laser, and $k$ is the imaginary part of the complex index of refraction $n^* = n + ik$, with $n = \sqrt{(|\epsilon^*| + \epsilon_1)/2}$, $k = \sqrt{(|\epsilon^*| - \epsilon_1)/2}$, and $|\epsilon^*| = \sqrt{\epsilon_1^2 + \epsilon_2^2}$. The penetration depth of \SI{800}{\nm} and \SI{400}{\nm} in BTO, $\delta_\mathrm{BTO}^{\SI{800}{\nm}} = \SI{944}{\nm}$ and $\delta_\mathrm{BTO}^{\SI{400}{\nm}} = \SI{167}{\nm}$, are obtained from the equations above, calculating the complex dielectric function as $\epsilon^* = \epsilon_1 + i \epsilon_2 = \sin{\theta_i}^2[1+\tan{\theta_i}^2(\frac{1-\rho}{1+\rho})^2]$, where $\rho = \tan{\Psi}\exp(i\Delta)$, with $\Psi$ and $\Delta$ reported in Ref. \cite{chernova_strain-controlled_2015}. Due to the large penetration depths $\delta_\mathrm{BTO}^{\SI{800}{\nm}}$ and $\delta_\mathrm{BTO}^{\SI{400}{\nm}}$, we probe the entire BTO thickness ($d_\mathrm{BTO} = \SI{34.5}{\nm}$).

\clearpage

\begin{table*}[h]
\caption[Physical constants]{\label{SI:tab:constants}\textbf{Physical constants used in the calculation of two-temperature model and strain model.} The film thicknesses $d$ result from $\theta$-$2\theta$ measurements (\ref{SI:fig:theta2theta}). The penetration depth $\delta$ and the reflectivity $R_p$ are discussed in \ref{si:note:transmittance_profile} and \ref{si:note:penetration_depths}. The volumetric heat capacities $C_e$ and $C_p$ of SRO and GSO are calculated from Refs. \cite{yamanaka_thermophysical_2004} and \cite{hidde_thermal_2018}, respectively, while those of BTO are discussed in \ref{si:note:2tm}. The mass density $\rho$, the bulk modulus $B$, and the Poisson ratio $\nu$ are taken from Ref. \cite{de_jong_charting_2015}. The longitudinal sound velocity $v = \sqrt{3B(1-\nu)/[\rho(1+\nu)]}$ \cite{thomsen_surface_1986}. The linear expansion coefficient $\beta$ of SRO and GSO, and $\beta_{T>T_c}$ of BTO are obtained from data shown in \ref{SI:fig:c_vs_temperature}, while $\overline{\beta}_{T<T_c}$ of BTO is a fit parameter, as discussed in \ref{si:note:strain_model}. The thermal conductivity $K_p$ and diffusivity $D_p = K_p / C_p$ are given in the respective references reported in the Table below. The acoustic reflection coefficient is given by $R_Z = (Z_2 - Z_1)/(Z_2 + Z_1)$, where $Z_1$ and $Z_2$ are the acoustic impedances at the two sides of the interface, with $Z = \rho v$ and the strain wave travelling from material 1 to 2. For example, going from BTO to SRO, $R_Z = 0.13$, while going from BTO to the vacuum interface, $R_Z = -1$.}
\begin{tabular}{ccccc}
 constants & unit & BTO & SRO & GSO \\ 
 \hline
 $d$ & \SI{}{\nm} & $34.5$ & $47$ & \SI{0.5e6}{} \\ 
 $\delta$ & \SI{}{\nm} & $17.9$ & $21.5$ & $26.1$ \\
 $R_p$ & - & $0.178$ & $0.022$ & $0.018$ \\
 $C_p$ & \SI{}{\joule\per\cubic\meter\per\K} & \SI{2.8e6}{} & \SI{2.8e6}{} \cite{yamanaka_thermophysical_2004} & \SI{2.1e6}{} \cite{hidde_thermal_2018}\\
 $C_e$ & \SI{}{\joule\per\cubic\meter\per\K} & \SI{4.2e4}{} & \SI{2.3e5}{} \cite{yamanaka_thermophysical_2004} & \SI{9e4}{} \cite{hidde_thermal_2018} \\
 $\rho$ \cite{de_jong_charting_2015} & \SI{}{\g\per\cubic\cm} & $5.93$ & $6.46$ & $6.6$ \\ 
 $m$ \cite{mills_quantities_1993} & \SI{}{\g\per\mole} & $233.19$ & $236.69$ & $250.2$ \\ 
 $\beta$ & \SI{}{\per\K} & \SI{1.33e-5}{} ($\beta_{T>T_c}$) & \SI{1.58e-5}{} & \SI{0.76e-5}{} \\  
 $K_p$ & \SI{}{\watt\per\m\per\K} & \SI{2.73}{} \cite{he_heat_2004} & \SI{5.72}{} \cite{yamanaka_thermophysical_2004} & \SI{2.53}{} \cite{hidde_thermal_2018} \\  
 $D_p$ & \SI{}{\square\mm\per\s} & \SI{1.11}{} \cite{he_heat_2004} & \SI{2.05}{} \cite{yamanaka_thermophysical_2004} & \SI{1.22}{} \cite{hidde_thermal_2018} \\  
 $B$ \cite{de_jong_charting_2015} & \SI{}{\giga\pascal} & $107$ & $166$ & $163$ \\
 $\nu$ \cite{de_jong_charting_2015} & - & 0.3 & $0.31$ & $0.29$ \\
 $v$ \cite{thomsen_surface_1986} & \SI{}{\m\per\s} & $5399$ & $6372$ & $6386$ \\
 $Z$ & \SI{}{\kg\per\square\m\per\s} & \SI{3.2e7}{} & \SI{4.1e7}{} & \SI{4.2e7}{} \\
 $R_Z$ & - & \SI{-1}{} & \SI{0.13}{} & \SI{0.01}{} \\ 
\hline
\end{tabular}
\end{table*}

\clearpage

\section{Two-temperature model}\label{si:note:2tm}

Electron and phonon temperatures, $T_e(z,t)$ and $T_p(z,t)$, result from the analytical solution of the two-temperature model (2TM), consisting of the following coupled equations \cite{anisimov_electron_1974}:
\begin{equation}\label{eq:2TM}
\begin{aligned}
  C_e\frac{\partial T_e}{\partial t} &= -g(T_e - T_p) + S(z,t),\\
  C_p\frac{\partial T_p}{\partial t} &= g(T_e - T_p),
\end{aligned}
\end{equation}
where $C_e$ and $C_p$ are volumetric heat capacities, respectively, and $g$ is the electron-phonon coupling. The source term $S(z, t)$ represents the absorbed power density (\SI{}{\watt\per\cubic\meter}) of the pump laser in our sample as a function of $t$ and $z$. It is defined as $S(z,t) = F_\mathrm{abs}\exp{(-z/(\delta\cos{\theta_t}))}\exp{(-t^2/(2\tau_\mathrm{OL}^2))} / (\delta\tau_\mathrm{OL})$, where $F_\mathrm{abs}$ is the pump absorbed fluence (\ref{si:note:transmittance_profile}), $\delta$ is the penetration depth (\ref{si:note:penetration_depths}), $\theta_t$ is the angle of the transmitted optical beam with respect to the surface normal, and $\tau_\mathrm{OL}$ is the optical laser pulse duration (\ref{SI:note:BAM}).  

In general, equation \eqref{eq:2TM} contains also the diffusion terms $K_e \partial^2 T_e / \partial z^2$ and $K_p \partial^2 T_p / \partial z^2$, where $K_e$ and $K_p$ are carrier and thermal conductivity, respectively. In the particular case of BTO, carrier and thermal diffusion terms can be neglected because the parameters $D_e/(\nu\delta_\mathrm{BTO})=\SI{2.5e-5}{}$ and $D_p/(\nu\delta_\mathrm{BTO})=\SI{1.1e-2}{}$ are relatively small. Here, the electron diffusion coefficient is calculated from the Einstein relation $D_e = \mu k_B T / q = \SI{2.6e-7}{\square\meter\per\s}$, where $\mu = \SI{0.1}{\square\cm\per\volt\per\s}$ is the BTO electron mobility \cite{muller_ambipolar_1989}, $k_B$ is the Boltzmann constant, $T$ is the sample temperature and $q$ is the electron charge, while the BTO thermal diffusivity \cite{he_heat_2004} is $D_p = \SI{1.11}{\square\mm\per\s}$.

We turn now to discuss the volumetric heat capacities $C_e$ and $C_p$. Electron and phonon volumetric heat capacities of SRO and GSO are taken from Refs. \cite{yamanaka_thermophysical_2004, hidde_thermal_2018}, assuming $T=\SI{300}{\K}$. This assumption is justified by the fact that in GSO the transmittance of the \SI{266}{\nm} beam is essentially $0\%$, while in the SRO thin film only $12\%$ of the incident fluence is absorbed. For the absorbed fluence $F_\mathrm{abs} = \SI{0.33}{\mJ\per\square\cm}$ (\ref{si:note:transmittance_profile}), the maximum temperature increase \cite{thomsen_surface_1986} in SRO is $\Delta T = F_\mathrm{abs}^\mathrm{SRO}/(\delta_\mathrm{SRO} C_p^\mathrm{SRO}) = \SI{55}{\K}$, with a consequent increase of $C_p^\mathrm{SRO}$ of only $2\%$. Conversely, $C_e^\mathrm{SRO}$ increases by $18\%$, but the absolute value remains one order of magnitude smaller than $C_p^\mathrm{SRO}$. The volumetric heat capacity $C_p^\mathrm{BTO}$ as a function of $T$ is reported in Ref. \cite{wang_strain_2021}. Given the weak temperature dependence for $T>\SI{300}{\K}$, we consider $C_p^\mathrm{BTO} = \SI{2.8e6}{\joule\per\cubic\meter\per\K}$ and calculate $\Delta T = F_\mathrm{abs}^\mathrm{BTO}/(\delta_\mathrm{BTO} C_p^\mathrm{BTO}) = \SI{375}{\K}$ and $\SI{196}{\K}$, for $F_\mathrm{abs}^\mathrm{BTO} = \SI{1.88}{\mJ\per\square\cm}$ and $\SI{0.98}{\mJ\per\square\cm}$ (\ref{si:note:transmittance_profile}), respectively. Due to the weak dependence on $T$, the average of $C_p^\mathrm{BTO}$ between \SI{300}{\K} and $\SI{300}{\K} + \Delta T$ yields the self-consistent result of $C_p^\mathrm{BTO} = \SI{2.8e6}{\joule\per\cubic\meter\per\K}$ \cite{wang_strain_2021}. Finally, to estimate $C_e^\mathrm{BTO}$, we rely on the available data from other titanates, e.g. SrTiO$_3$ \cite{de_ligny_high-temperature_1996} and CaTiO$_3$ \cite{guyot_high-temperature_1993}, and observe that they have an average ratio $C_e/C_p = 0.015$ for the $T$ ranges considered here. This yields $C_e^\mathrm{BTO} = \SI{4.2e4}{\joule\per\cubic\meter\per\K}$. Although $C_e^\mathrm{BTO}$ is a temperature-dependent quantity, for simplicity, we assume it to be constant here. For completeness, all the physical constants employed in the solution of the 2TM are reported in \ref{SI:tab:constants}. 

We focus now on the analytical solution of the 2TM. The second equation of equation \eqref{eq:2TM} can be written as:
\begin{equation}\label{SI:eq:2TM_e}
  T_e = \frac{C_p}{g}\frac{\partial T_p}{\partial t} + T_p,
\end{equation}
which is then inserted in the first equation of \eqref{eq:2TM} and yields the following second order differential equation of $T_p$:
\begin{equation}\label{SI:eq:2TM_p}
  \frac{C_eC_p}{g}\frac{\partial^2 T_p}{\partial t^2} + (C_e + C_p) \frac{\partial T_p}{\partial t} = S(z,t).
\end{equation}
Given the following initial and boundary conditions: $T_e(z,0)=T_p(z,0)=T_0=\SI{300}{\K}$, $\partial T_p / \partial z (0,t)=0$, $\partial T_p / \partial t (z,0)=0$, we can derive the lattice temperature $T_p(z,t)$ as an analytical solution of equation \eqref{SI:eq:2TM_p}:
\begin{equation}\label{SI:eq:Tp}
\begin{split}
  T_p(z,t) = T_0 + \frac{\sqrt{\pi}\mathcal{C}\tau_\mathrm{OL}}{\sqrt{2}\mathcal{B}}\exp{\left(\frac{\tau_\mathrm{OL}^2\mathcal{B}^2}{2\mathcal{A}^2}\right)}\operatorname{erf}\left(\frac{\mathcal{B}\tau_\mathrm{OL}}{\sqrt{2}A} - \frac{t}{\sqrt{2}\tau_\mathrm{OL}}\right)\exp{\left(-\frac{z}{\delta}\right)}\exp{\left(-\frac{\mathcal{B}}{\mathcal{A}}t\right)} \\
               + \frac{\sqrt{\pi}\mathcal{C}\tau_\mathrm{OL}}{\sqrt{2}\mathcal{B}}\exp{\left(-\frac{z}{\delta}\right)}\operatorname{erf}\left(\frac{1}{\sqrt{2}\tau_\mathrm{OL}}t\right) \\
               - \frac{\sqrt{\pi}\mathcal{C} \tau_\mathrm{OL}}{\sqrt{2}\mathcal{B}}\exp{\left(\frac{\tau_\mathrm{OL}^2\mathcal{B}^2}{2\mathcal{A}^2}\right)}\operatorname{erf}\left(\frac{\mathcal{B}\tau_\mathrm{OL}}{\sqrt{2}A}\right)\exp{\left(-\frac{z}{\delta}\right)}\exp{\left(-\frac{\mathcal{B}}{\mathcal{A}}t\right)},
\end{split}
\end{equation}
where $\mathcal{A} = C_e  C_p / g$, $\mathcal{B} = C_e + C_p$, and $\mathcal{C} = (1/\delta)F_\mathrm{abs}$. Substituting equation \eqref{SI:eq:Tp} in equation \eqref{SI:eq:2TM_e} we obtain the electron temperature $T_e(z,t)$:
\begin{equation}\label{SI:eq:Te}
\begin{split}
  T_e(z,t) = T_0 - \frac{\sqrt{\pi}\mathcal{C} \tau_\mathrm{OL}}{\sqrt{2}\mathcal{B}}\frac{C_p}{C_e}\exp{\left(\frac{\tau_\mathrm{OL}^2\mathcal{B}^2}{2\mathcal{A}^2}\right)}\operatorname{erf}\left(\frac{\mathcal{B}\tau_\mathrm{OL}}{\sqrt{2}A} - \frac{t}{\sqrt{2}\tau_\mathrm{OL}}\right)\exp{\left(-\frac{z}{\delta}\right)} \exp{\left(-\frac{\mathcal{B}}{\mathcal{A}}t\right)} \\
               + \frac{\sqrt{\pi}\mathcal{C} \tau_\mathrm{OL}}{\sqrt{2}\mathcal{B}}\exp{\left(-\frac{z}{\delta}\right)}\operatorname{erf}\left(\frac{1}{\sqrt{2}\tau_\mathrm{OL}}t\right)  \\
               + \frac{\sqrt{\pi}\mathcal{C} \tau_\mathrm{OL}}{\sqrt{2}\mathcal{B}}\frac{C_p}{C_e}\exp{\left(\frac{\tau_\mathrm{OL}^2\mathcal{B}^2}{2\mathcal{A}^2}\right)}\operatorname{erf}\left(\frac{\mathcal{B}\tau_\mathrm{OL}}{\sqrt{2}\mathcal{A}}\right)\exp{\left(-\frac{z}{\delta}\right)} \exp{\left(-\frac{\mathcal{B}}{\mathcal{A}}t\right)}.
\end{split}
\end{equation}

\begin{figure*}[h]
\centering
\includegraphics[width=1\textwidth]
{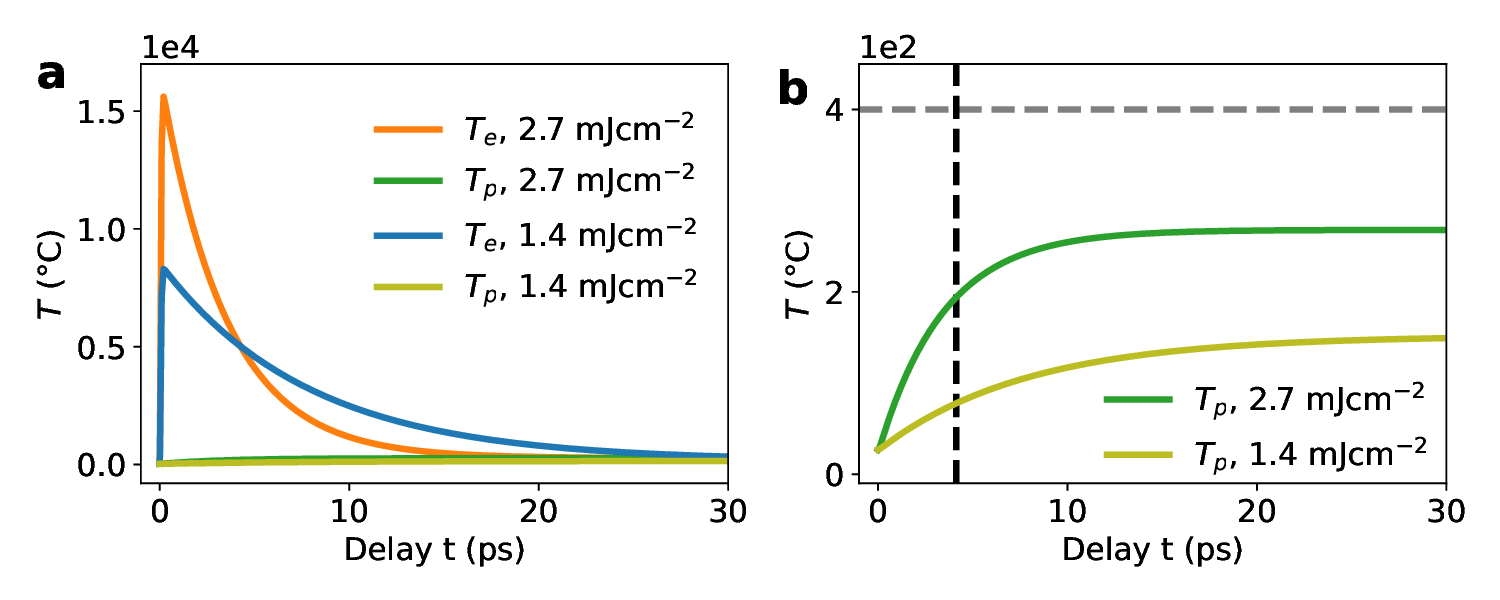}
\caption[Electron and lattice temperature as a function of delay $t$]{\textbf{Electron and lattice temperature as a function of delay $t$.} (a) $T_e$ and $T_p$ averaged over $d_\mathrm{BTO}$ as a function of delay $t$ for the incident pump fluences $F_\mathrm{in} = \SI{1.4}{\milli\joule\per\square\cm}$ and $F_\mathrm{in} = \SI{2.7}{\milli\joule\per\square\cm}$, calculated as detailed in \ref{si:note:2tm}. (b) the same as panel (a) with focus on the temperature range of $T_p$. The gray dashed line indicates $T_c = \SI{400}{\celsius}$ (\ref{SI:fig:c_vs_temperature}).}
\centering
\label{si:fig:Temperature_profile}
\end{figure*}

The dependence of $T_e$ and $T_p$ (averaged over $d_\mathrm{BTO}$) on the delay $t$ is reported in \ref{si:fig:Temperature_profile} for the two incident pump fluences $F_\mathrm{in}=\SI{1.4}{\milli\joule\per\square\cm}$ and $\SI{2.7}{\milli\joule\per\square\cm}$ and the corresponding fit parameter $g$ (\ref{si:tab:fit_results}), while \ref{SI:fig:T_strain_XRD_sim_1_6_mJ}a and \ref{SI:fig:T_strain_XRD_sim_2_7_mJ}a report the 2D maps of $T_p(z,t)$ as a function of delay $t$ and depth $z$ at the respective fluences.

\clearpage

\section{Strain model}\label{si:note:strain_model}

The total stress experienced by our sample upon optical laser excitation can be written as \cite{wright_acoustic_1995, ruello_physical_2015}:
\begin{equation}\label{eq:stress}
    \sigma = \rho v^2 \eta + \sigma_{DP}\left(T_e,\frac{\partial E_g}{\partial p}\right) + \sigma_{TE}(T_p,\beta),
\end{equation}
where $Y = \rho v^2$  is the Young’s modulus, $\rho$ is the mass density and $v$ is the longitudinal speed of sound (\ref{SI:tab:constants}). The first term in equation \eqref{eq:stress} indicates the direct relationship between the stress $\sigma$, a force that induces a deformation of the material, and the strain $\eta$ that represents the resulting deformation of the material. The stress $\sigma$ causes the generation of a strain wave that propagates through the material and across the interface to the layer below. In equation \eqref{eq:stress}, $\partial E_g / \partial p$ indicates the variation of the bandgap as a function of the electronic pressure, and $\beta$ is the linear expansion coefficient.

Equation \eqref{eq:stress} can be written more explicitly as \cite{wright_acoustic_1995}:
\begin{equation}\label{SI:eq:stress_explicit}
  \sigma(z,t) = \rho v^2 \eta(z,t) - n_e B\frac{\partial E_g}{\partial p} - 3B\beta(T_p(z,t) - T_0),
\end{equation}
where $n_e = C_e(T_e(z,t) - T_0)/(E - E_g)$ is the carrier density \cite{ruello_physical_2015} (with unit \SI{}{\per\cubic\meter}), $B$ is the bulk modulus (\ref{SI:tab:constants}), $T_0=\SI{300}{\K}$ is the sample temperature at equilibrium, and $n_e (E - E_g)$ is the total energy density transferred from the optical photons to the electronic subsystem \cite{ruello_physical_2015}. We consider here only the dependence of $T_e$, $T_p$ and $\eta$ on the $z$ direction (sample depth). This one-dimensional approximation is justified by the large ratio between the laser excited area and the BTO thickness, leading to the film contraction/expansion only along the surface normal on the few tens of picoseconds timescale.

The relation between the stress $\sigma$ and the atomic displacement $u$ is described by the following one-dimensional lattice strain wave equation \cite{thomsen_surface_1986}:
\begin{equation}\label{eq:strain}
    \rho \frac{\partial^2 u(z,t)}{\partial t^2} = \frac{\partial \sigma(z,t)}{\partial z},
\end{equation}
which can be recast as a function of $\eta(z,t) = \partial u(z,t)/\partial z$: 
\begin{equation}\label{SI:eq:strain_stress}
  \frac{\partial^2 \eta(z,t)}{\partial t^2} = \frac{1}{\rho}\frac{\partial^2 \sigma(z,t)}{\partial z^2}.
\end{equation}

Substituting equation \eqref{SI:eq:stress_explicit} in equation 
\eqref{SI:eq:strain_stress} yields:
\begin{equation}\label{SI:eq:strain_diff}
  \frac{\partial^2 \eta(z,t)}{\partial t^2}=  v^2 \frac{\partial^2 \eta(z,t)}{\partial z^2} - \frac{C_e}{\rho(E - E_g)\delta^2}B\frac{\partial E_g}{\partial p}(T_e(z,t)-T_0) - \frac{3B\beta}{\rho\delta^2}(T_p(z,t)-T_0).
\end{equation}
Given the initial conditions \cite{thomsen_surface_1986} $\eta(z,0) = 0$, $\partial \eta/ \partial t (z,0)=0$, $\sigma(0,t)=0$, equation \eqref{SI:eq:strain_diff} provides the out-of-plane strain profile $\eta(z,t)$ by solving the following analytical integrals \cite{vladimirov_collection_1986}:
\begin{equation}\label{SI:eq:strain_integrals}
  \eta(z,t) =
    \begin{cases}
      \frac{1}{2v}\int_{0}^{t} \int_{z - v(t - \tau)}^{z + v(t - \tau)}\mathfrak{F}(\xi,\tau)d\xi d\tau, & \text{for $t < \frac{z}{v}$}\\
      \frac{1}{2v}\int_{0}^{t-\frac{z}{v}} \int_{v(t - \tau)-z}^{z + v(t - \tau)}\mathfrak{F}(\xi,\tau)d\xi d\tau \\ + \frac{1}{2v}\int_{t-\frac{z}{v}}^{t} \int_{z-v(t - \tau)}^{z + v(t - \tau)}\mathfrak{F}(\xi,\tau)d\xi d\tau + \mathfrak{F}(0,t-z/v)\frac{\delta^2}{v^2}, & \text{for $t > \frac{z}{v}$},
    \end{cases}       
\end{equation}
where $\mathfrak{F}(z,t) = - C_eB\frac{\partial E_g}{\partial p}(T_e(z,t)-T_0)/[\rho(E - E_g)\delta^2] - 3B\beta (T_p(z,t)-T_0)/(\rho \delta^2)$. Electron and phonon temperatures, $T_e(z,t)$ and $T_p(z,t)$, are determined using the 2TM (\ref{si:note:2tm}).

\begin{figure*}[!t]
\centering
\includegraphics[width=0.9\textwidth]
{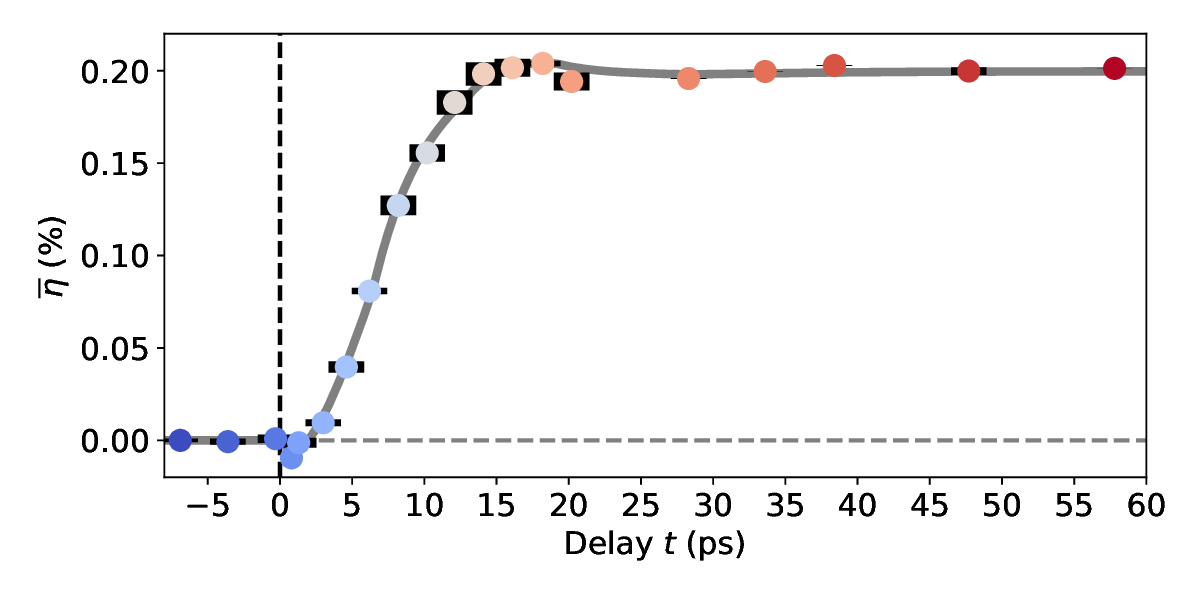}
\caption[$\overline{\eta}(t)$ with pump fluence $F_\mathrm{in} = \SI{1.4}{\milli\joule\per\square\cm}$]{\textbf{$\overline{\eta}(t)$ with pump fluence $F_\mathrm{in} = \SI{1.4}{\milli\joule\per\square\cm}$.} Average BTO out-of-plane strain $\overline{\eta}(t)$ as a function of pump-probe delay $t$, measured with the incident pump fluence $F_\mathrm{in} = \SI{1.4}{\milli\joule\per\square\cm}$. The error bars follow from the determination of $c$ from $I_\mathrm{XRD}(E_\nu) \pm \mathrm{SD}$, where $\mathrm{SD}$ is the standard deviation. The solid gray line is a fit to the data.}
\centering
\label{SI:fig:eta_vs_delay_at_1_4mJ}
\end{figure*}

Since the optical pump laser is absorbed mostly in the BTO thin film, but partially also in the SRO layer, there will be two strain waves originating at the vacuum/BTO and the BTO/SRO interfaces. These strain waves propagate through the sample and reflect at each interface, with acoustic reflection coefficients $R_Z$ (\ref{SI:tab:constants}). The almost identical acoustic impedance of SRO and GSO, leads to a $R_Z = 1\%$ at the SRO/GSO interface. Conversely, the strain wave undergoes a $13\%$ reflection at the BTO/SRO interface and $100\%$ reflection at the interface with vacuum, given the acoustic impedance $Z_\mathrm{vacuum} = 0$ (\ref{SI:tab:constants}). The superposition of the generated and reflected strain waves yields a total strain wave $\eta(z,t)$, which is then averaged over the BTO film thickness to obtain $\overline{\eta}(t) =  \int_0^{d_{\mathrm{BTO}}}\eta(z,t)dz/d_{\mathrm{BTO}}$. This quantity is used to fit the experimental average strain data obtained from tr-XRD measurements for $F_\mathrm{in} = \SI{2.7}{\milli\joule\per\square\cm}$ (Figure \ref{fig:Figure_1}d) and for $F_\mathrm{in} = \SI{1.4}{\milli\joule\per\square\cm}$ (\ref{SI:fig:eta_vs_delay_at_1_4mJ}), with the resulting fit parameters ($\partial E_g / \partial p$, $\overline{\beta}_{T<T_c}$, and $g$) reported in \ref{si:tab:fit_results}. The parameter $\partial E_g / \partial p$ has been discussed in the main text, we focus here on the discussion of $\overline{\beta}_{T<T_c}$, and $g$ fit results, while the calculations based on our strain model are presented in \ref{si:note:strain_model}.

\begin{table*}[h]
\caption[Strain model fit results]{\textbf{Strain model fit results.} Fit results of the experimental $\overline{\eta}(t)$ curves in Figure \ref{fig:Figure_1}d and \ref{SI:fig:eta_vs_delay_at_1_4mJ}, including $\partial E_g/\partial p$, the linear expansion coefficient $\overline{\beta}_{T<T_c}$ and the electron-phonon coupling $g$, measured at two different incident pump fluences $F_\mathrm{in}$.}
\begin{tabular}{ccccc}
\hline
 $F_\mathrm{in}$ (\SI{}{\milli\joule\per\square\cm}) & $\partial E_g/\partial p$ (\SI{}{\joule\per\pascal}) & $\overline{\beta}_{T<T_c}$ (\SI{}{\per\K}) & g (\SI{}{\watt\per\cubic\meter\per\K})  \\ 
\hline
  1.4 & \SI{-3.0(4)e-31}{} & \SI{8.5(1)e-6}{} & \SI{6.5(2)e15}{}  \\
\hline
  2.7 & \SI{-3.0(5)e-31}{} & \SI{2.5(2)e-6}{} & \SI{1.2(1)e16}{}  \\
\hline
\end{tabular}\label{si:tab:fit_results}
\end{table*}

\begin{figure*}[h]
\centering
\includegraphics[width=1\textwidth]
{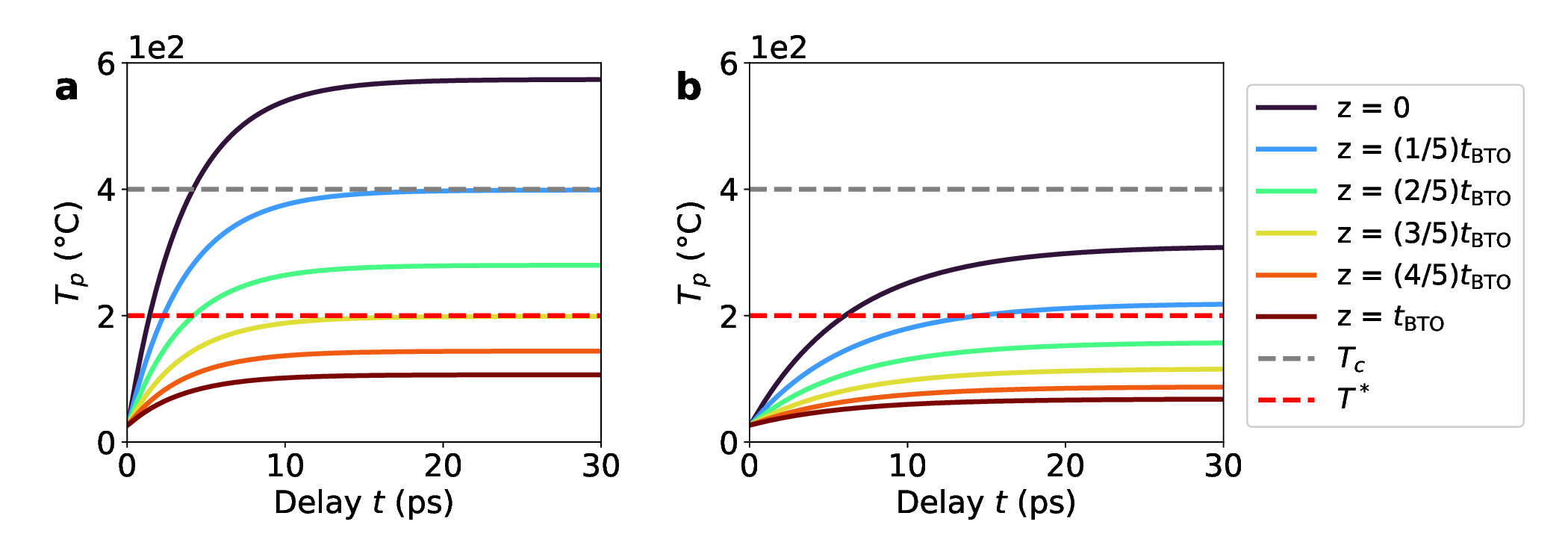}
\caption[Lattice temperature as a function of delay and depth]{\textbf{Lattice temperature as a function of delay and depth.} $T_p$ at different depths $z$ from the surface ($z=0$) as a function of delay $t$ for $F_\mathrm{in} = \SI{2.7}{\mJ\per\square\cm}$ (a) and $F_\mathrm{in} = \SI{1.4}{\mJ\per\square\cm}$ (b). The dashed gray[red] line indicates $T_c$ [$T^*$].}
\centering
\label{SI:fig:Tp_at_z_vs_delay}
\end{figure*}

The linear expansion coefficient $\overline{\beta}_{T<T_c}$ is the fit parameter referring to the average $\beta$ in the range $T_0<T<T_c$, with Curie temperature $T_c=\SI{400}{\celsius}$. While for $T>T_c$, the linear expansion coefficient $\beta$ is constant and positive ($\beta_{T>T_c} = \SI{1.33e-5}{\per\kelvin}$), it is positive for $T_0<T<T^*$, and negative for $T^*<T<T_c$ (\ref{SI:fig:c_vs_temperature}). In our sample, the temperature below $T_c$ at which $\beta$ changes sign is $T^*=\SI{200}{\celsius}$ (\ref{SI:fig:c_vs_temperature}). Based on our model, in the metastable state ($t>\SI{20}{\ps}$), only $\approx \SI{7}{\nm}$ of the BTO film is at $T^*<T<T_c$ for $F_\mathrm{in} = \SI{1.4}{\milli\joule\per\square\cm}$, while $\approx \SI{14}{\nm}$ of the BTO film is at $T^*<T<T_c$ for $F_\mathrm{in} = \SI{2.7}{\milli\joule\per\square\cm}$ (\ref{SI:fig:Tp_at_z_vs_delay}). As a result, in the latter case the portion of the sample at $\beta<0$ is larger than in the former case, thus $\overline{\beta}_{T<T_c}$ is expected to be smaller, as shown by our fit results (\ref{si:tab:fit_results}). At the same time, for both fluences, the larger portion of the sample ($\approx 14-27 \SI{}{\nm}$) is at $T_0<T<T^*$, hence $\overline{\beta}_{T<T_c}$ is expected to be positive, as obtained by our fit results.

\begin{figure*}[h]
\centering
\includegraphics[width=0.6\textwidth]
{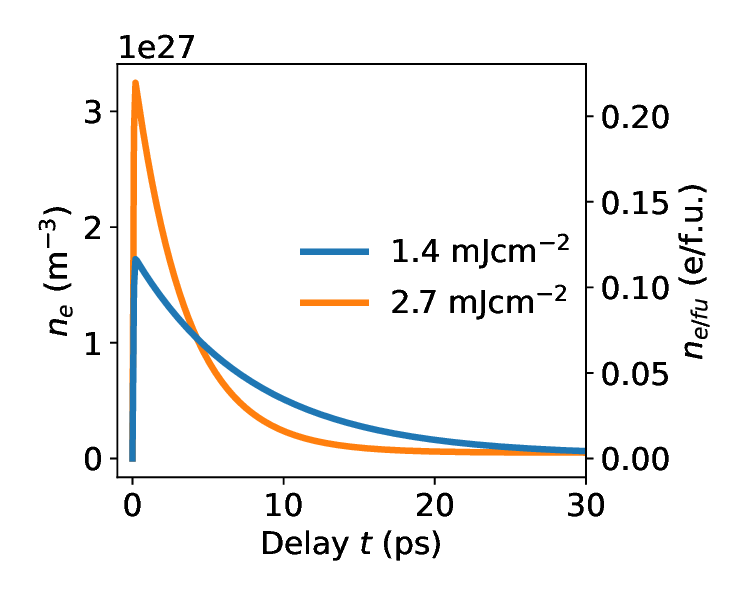}
\caption[Electron density as a function of delay and fluence]{\textbf{Electron density as a function of delay.} Electron density calculated as $n_e = C_e(T_e(t) - T_0)/(E-E_g)$ \cite{ruello_physical_2015} for two different pump fluences, and corresponding concentration of photoexcited electrons per formula unit (e/f.u.) $n_{e/fu}$, where $n_{e/fu} = n_e V$ with the formula unit volume $V = c_0^3 = \SI{67.62e-30}{\cubic\metre}$. Here, $T_e(t)$ refers to the electron temperature averaged over $d_\mathrm{BTO}$.}
\centering
\label{SI:fig:electron_density}
\end{figure*}

The electron-phonon coupling $g$ of the 2TM (\ref{si:note:2tm}) is a temperature dependent quantity, however for simplicity in our model it is assumed to be constant, similarly to most current investigations of short-pulse laser excitation \cite{lin_electron-phonon_2008}. In general, the dependence of $g$ on $T_e$ is related to the electron density of states, and, if the $d$ band electrons are below the Fermi level without crossing it (as in BTO), $g$ increases with $T_e$. In particular, within the free electron gas model \cite{lin_electron-phonon_2008}, $g$ is linearly dependent on the electron density $n_e$. This approximation is expected to be valid for relatively low electron temperatures \cite{wang_time-resolved_1994}, as in our case. In fact, the resulting $g$ fit parameters (\ref{si:tab:fit_results}) scale approximately as the peak electron density given by $n_{e,max} = C_e(T_{e,max} - T_0)/(E-E_g)$ \cite{ruello_physical_2015} (\ref{SI:fig:electron_density}). Specifically, $n_{e,max} = \SI{1.7e+27}{\per\cubic\meter}$ and \SI{3.2e+27}{\per\cubic\meter} for $F_\mathrm{in}=\SI{1.4}{\milli\joule\per\square\cm}$ and $\SI{2.7}{\milli\joule\per\square\cm}$, respectively. For comparison, the electron density in copper, with one electron per atom in the conduction band ($N_e = 1$), is $n_e^\mathrm{Cu} = N_e \rho  N_A / M = \SI{8.5e28}{\per\cubic\m}$, where $\rho=\SI{8.96}{\gram\per\cubic\cm}$ is the mass density \cite{crc_handbook}, $N_A=\SI{6.022e23}{\text{atoms}\per\mol}$ is the Avogadro's number \cite{nist_constants}, and $M=\SI{63.55}{\gram\per\mol}$ is the atomic mass \cite{crc_handbook}. The absolute value of the fit $g$ parameters are of similar order of magnitude as for $\mathrm{SrRuO_3}$ \cite{wang_coupling_2019} and other perovskites \cite{chan_uncovering_2021}. Moreover, it is worth noting that the obtained electron densities and the fit $g$ values are more than one order of magnitude smaller than the corresponding parameters in metals \cite{elsayed-ali_time-resolved_1987, hohlfeld_electron_2000, lin_electron-phonon_2008}, e.g., the electron-phonon coupling in Cu \cite{lin_electron-phonon_2008} is $g\approx\SI{5e+17}{\watt\per\cubic\meter\per\K}$.

\clearpage

\section{Temperature, strain and diffraction curve calculations}\label{si:note:temp_strain_diffr_calculations}

\begin{figure}[!t]
\centering
\includegraphics[width=1\textwidth]{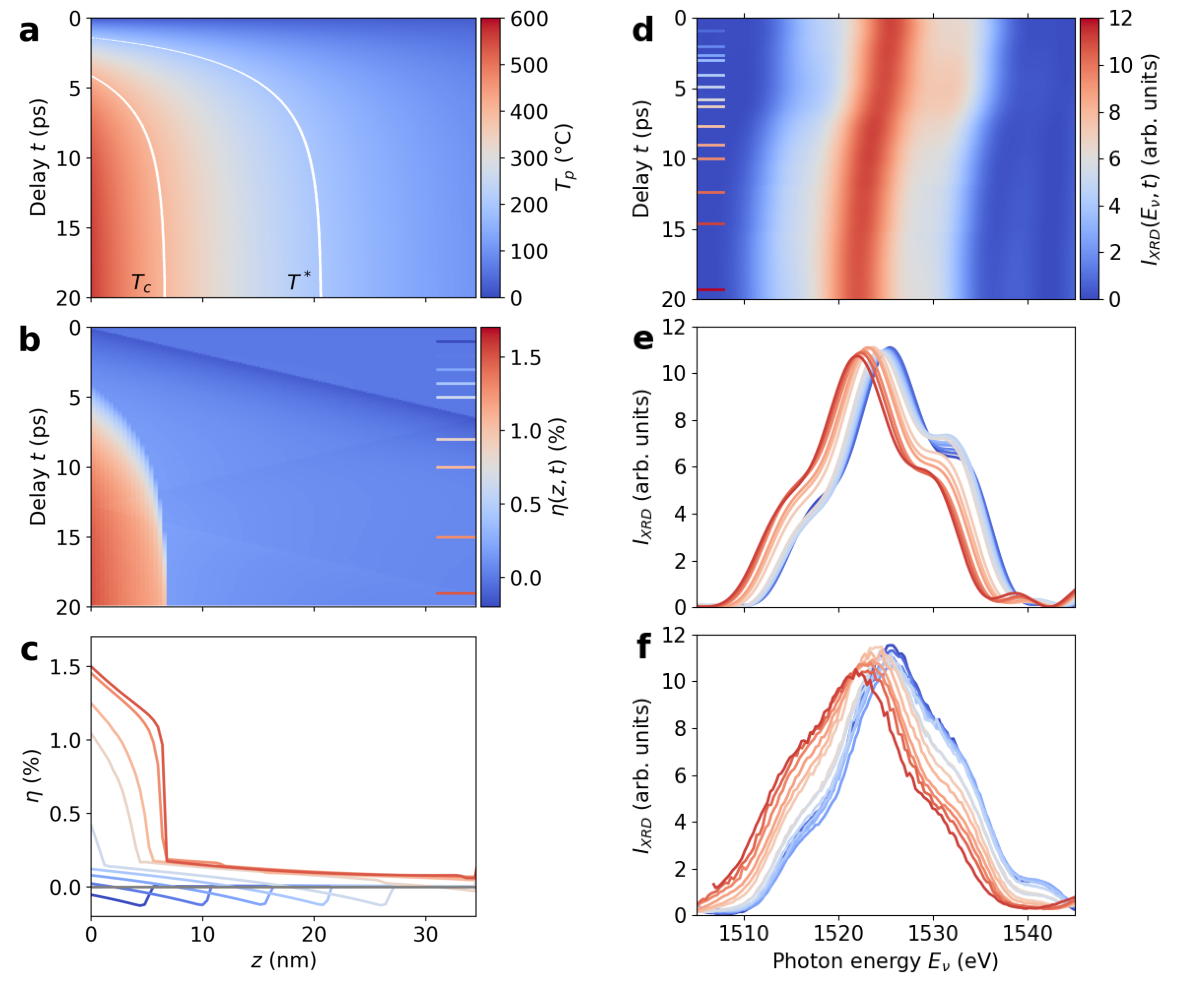}
\caption[Strain model calculations at $F_\mathrm{in} = \SI{2.7}{\mJ\per\square\cm}$]{\textbf{Temporal and spatial dependance of lattice temperature and strain, and diffraction curves at fixed delays, with $F_\mathrm{in} = \SI{2.7}{\mJ\per\square\cm}$.} (a) 2D map of the lattice temperature $T_p$ as a function of the delay $t$ and the distance $z$ from the BTO surface. The white lines indicate the Curie temperature $T_c = \SI{400}{\degreeCelsius}$ and $T^* = \SI{200}{\degreeCelsius}$ (\ref{SI:fig:c_vs_temperature}). (b) 2D map of the BTO strain $\eta(z,t)$ as a function of $t$ and $z$. (c) Strain profiles $\eta(z,t)$ at time delays $t$ marked in the panel (b). (d) 2D map of the simulated diffraction curves $I_\mathrm{XRD}(t)$ as a function of delay $t$ and photon energy $E_\nu$. (e) Simulated diffraction curves $I_\mathrm{XRD}(E_\nu)$ at time delays $t$ marked in panel (d). (f) Experimental diffraction curves at time delays $t$ marked in panel (d). The data shown here refer to the incident pump fluence $F_\mathrm{in} = \SI{2.7}{\mJ\per\square\cm}$ and results from the solution of the 2TM and the 1D strain wave equation (\ref{si:note:2tm} and \ref{si:note:strain_model}).}
\centering
\label{SI:fig:T_strain_XRD_sim_2_7_mJ}
\end{figure}

\begin{figure*}[!t]
\centering
\includegraphics[width=1\textwidth]
{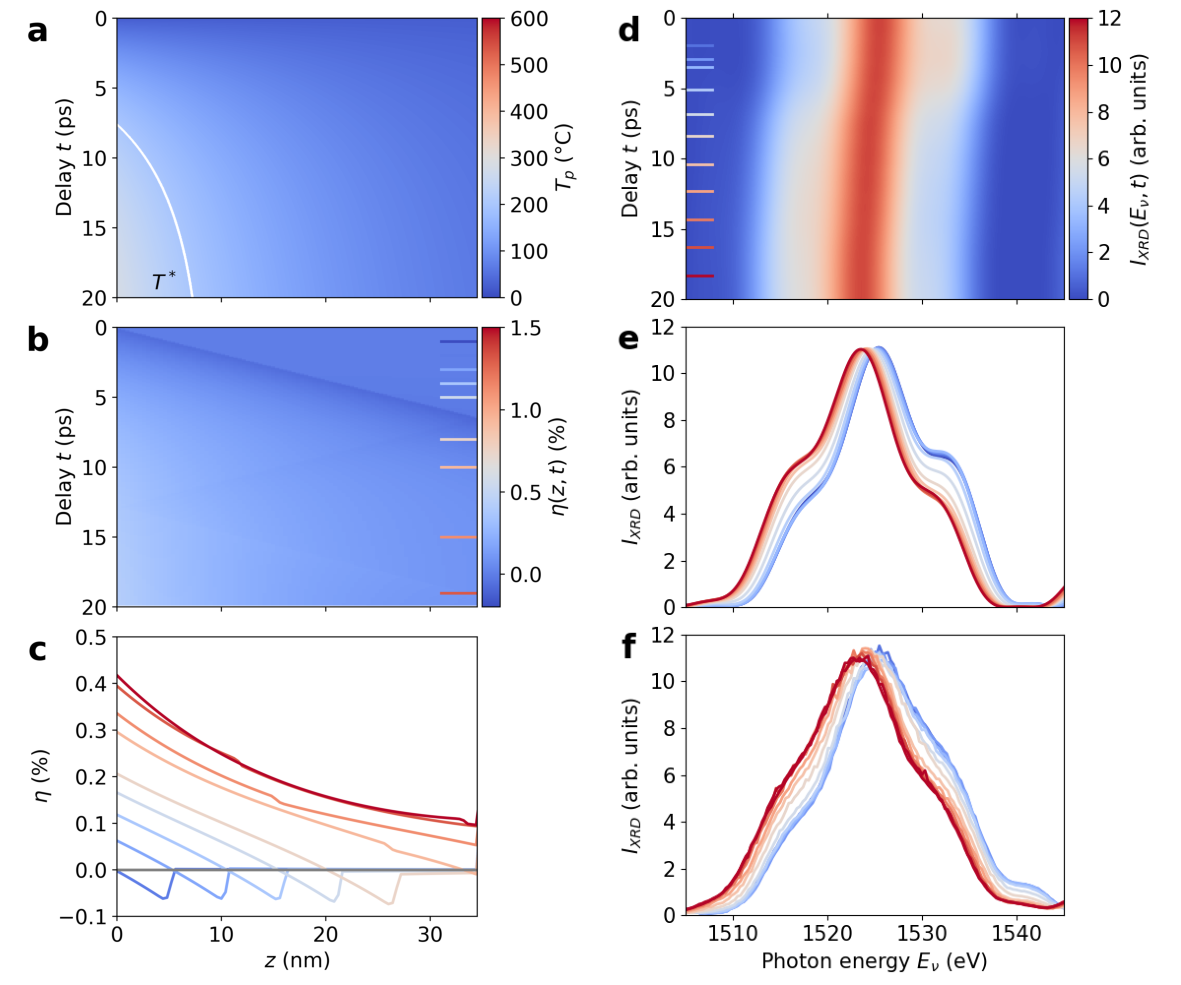}
\caption[Strain model calculations at $F_\mathrm{in} = \SI{1.4}{\mJ\per\square\cm}$]{\textbf{Temporal and spatial dependance of lattice temperature and strain, and diffraction curves at fixed delays, with $F_\mathrm{in} = \SI{1.4}{\mJ\per\square\cm}$.} (a) 2D map of the lattice temperature $T_p$ as a function of the delay $t$ and the distance $z$ from the BTO surface. The white line indicates the temperature $T^* = \SI{200}{\degreeCelsius}$ (\ref{SI:fig:c_vs_temperature}). (b) 2D map of the BTO strain $\eta(z,t)$ as a function of delay $t$ and $z$. (c) Strain profiles $\eta(z,t)$ at time delays $t$ marked in the panel (b). (d) 2D map of the simulated diffraction curves $I_\mathrm{XRD}(t)$ as a function of delay $t$ and photon energy $E_\nu$. (e) Selected diffraction curves $I_\mathrm{XRD}(E_\nu)$ at time delays $t$ marked in panel (d). (f) Experimental diffraction curves at time delays $t$ marked in panel (d). The data shown here refer to the incident pump fluence $F_\mathrm{in} = \SI{1.4}{\mJ\per\square\cm}$ and result from the solution of the 2TM and the 1D strain wave equation (\ref{si:note:2tm} and \ref{si:note:strain_model}).}
\centering
\label{SI:fig:T_strain_XRD_sim_1_6_mJ}
\end{figure*}

Given the fit parameters in \ref{si:tab:fit_results}, we calculate the lattice temperature $T_p(z,t)$ in the BTO thin film for $F_\mathrm{in} = \SI{2.7}{\mJ\per\square\cm}$ (\ref{SI:fig:T_strain_XRD_sim_2_7_mJ}a). Under this condition, $19\%$ of BTO reaches $T_c$ after \SI{20}{\ps}, while for $F_\mathrm{in} = \SI{1.4}{\mJ\per\square\cm}$ the sample remains always below $T_c$ (\ref{SI:fig:T_strain_XRD_sim_1_6_mJ}a). Hence, a higher pump fluence leads to a larger portion of the BTO reaching higher temperatures and in shorter time. We note that when the pump photon energy $E$ is below or $\lesssim\SI{0.5}{\eV}$ above the bandgap $E_g$, peak power intensities in the range from \SI{}{\kilo\watt\per\square\cm} to \SI{}{\giga\watt\per\square\cm} lead to a sample temperature increase below \SI{50}{\kelvin} \cite{daranciang_ultrafast_2012, schick_localized_2014, matzen_tuning_2019, ahn_dynamic_2021, ganguly_photostrictive_2024}. In our experiment, given the relatively high peak power intensity in the range \numrange[range-phrase = --]{20}{38} \SI{}{\giga\watt\per\square\cm}, and the pump photon energy $\approx \SI{1.2}{\eV}$ above the bandgap, sample heating is taken into account. 

\ref{SI:fig:T_strain_XRD_sim_2_7_mJ}b reports the strain map $\eta(z,t)$ in the BTO film based on the results of $T_p(z,t)$ and $T_e(z,t)$, with calculation details reported in \ref{si:note:strain_model}. Here, it can be clearly seen how the regions of the sample below [above] $T_c$ display smaller [larger] $\eta(z,t)$. In particular, for $t<\SI{3.1}{\ps}$ the average strain $\overline{\eta}(t)$ is negative, primarily due to the compressive strain from the deformation potential (Figure \ref{fig:Figure_1}d). Moreover, the three straight profiles in the \ref{SI:fig:T_strain_XRD_sim_2_7_mJ}b mark the front of the strain wave propagating in the BTO at the sound speed and then reflected at the interface with SRO and air.

\begin{figure*}[!t]
\centering
\includegraphics[width=1\textwidth]
{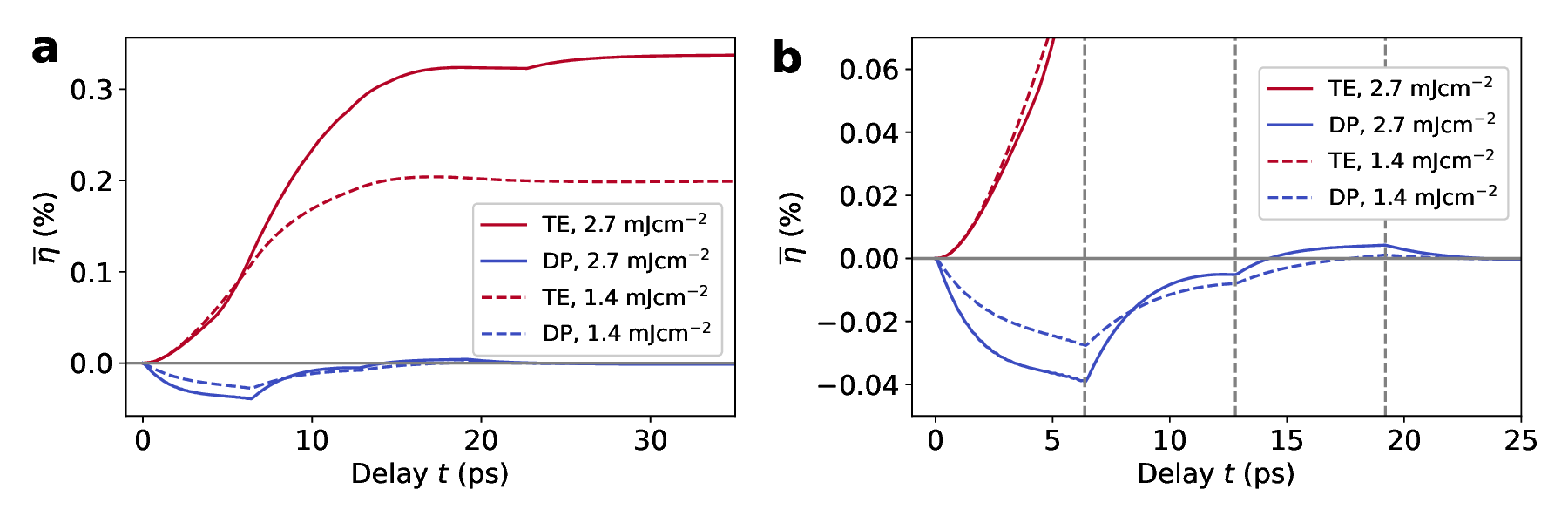}
\caption[Deformation potential and thermoelastic contributions]{\textbf{Deformation potential and thermoelastic contributions.} (a): Thermoelastic (TE, red lines) and deformation potential (DP, blue lines) strain contributions at $F_\mathrm{in} = \SI{2.7}{\mJ\per\square\cm}$ (solid lines) and $F_\mathrm{in} = \SI{1.4}{\mJ\per\square\cm}$ (dashed lines). (b): same plot as in panel (a) with focus on the strain range $-0.05\%<\overline{\eta}(t)<0.07\%$ and delay range $\SI{-1}{\ps}<t<\SI{25}{\ps}$. The vertical gray dashed lines at \SI{6.4}{\ps}, \SI{12.8}{\ps}, \SI{19.2}{\ps} mark discontinuities in the DP strain profile, which are due to reflections of the DP strain wave at the BTO/SRO, BTO/air, and again BTO/SRO interfaces (\ref{si:note:transmittance_profile}). The time constants result from the speed of sound in BTO (\ref{SI:tab:constants}) multiplied by the traveled distance, which is $d_\mathrm{BTO}$, $2d_\mathrm{BTO}$ and $3d_\mathrm{BTO}$, respectively. Discontinuities in the TE profile are too small to be seen because of the relatively large absolute value of the average TE strain that is mostly contributed by the BTO region near the surface (\ref{SI:fig:T_strain_XRD_sim_2_7_mJ}b).}
\centering
\label{si:fig:TE_DP_strain}
\end{figure*}

A few selected strain profiles at fixed delays are highlighted in \ref{SI:fig:T_strain_XRD_sim_2_7_mJ}c. Within the first \SI{15}{\ps} after $t = \SI{0}{\ps}$ (\ref{SI:note:BAM}) the strain profile $\eta(z,t)$ undergoes relatively large changes resulting from the varying $T_e$ and $T_p$, and the corresponding DP and TE contributions (\ref{si:fig:TE_DP_strain}). After $\approx \SI{15}{\ps}$ the strain profile reaches a metastable state at least for the following few tens of picoseconds. The maximum average tensile strain of this metastable state is directly proportional to the incident pump fluence and it depends on the film thickness (\ref{SI:fig:T_strain_XRD_sim_2_7_mJ}b-c and \ref{SI:fig:T_strain_XRD_sim_1_6_mJ}b-c).

\begin{figure*}[h]
\centering
\includegraphics[width=0.6\textwidth]
{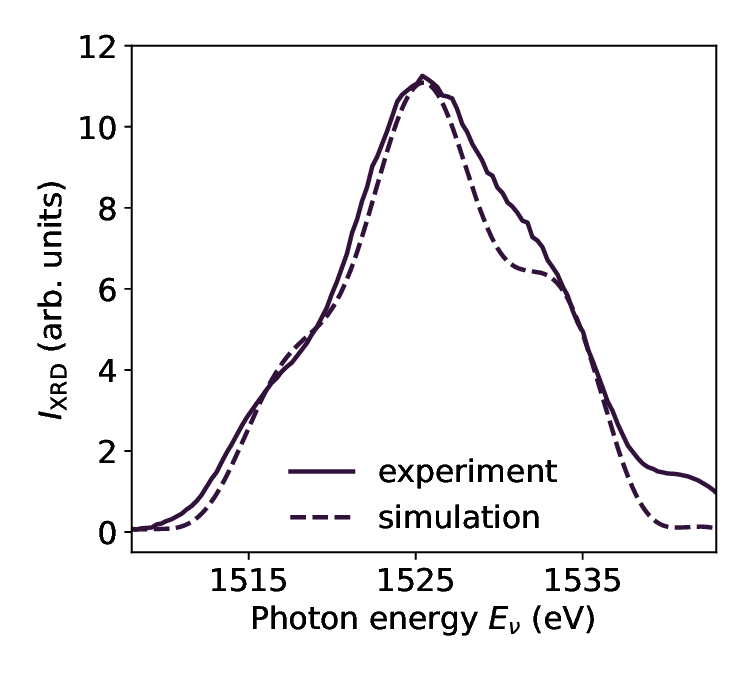}
\caption[Comparison of simulated and experimental $I_\mathrm{XRD}(E_\nu)$]{\textbf{Comparison of simulated and experimental $I_\mathrm{XRD}(E_\nu)$ at negative delay.} Comparison of simulated (dashed line) and experimental (solid line) diffraction curve before the arrival of the pump laser. The broadening of the diffraction curves is calculated as $c_\mathrm{std} = (\SI{1240000}{\eV\pm})/(2 E_{\nu,\mathrm{std}}\sin\theta)$, where $E_{\nu,\mathrm{std}}$ is the standard deviation of energy values, around the (001) BTO peak, weighted by $I_\mathrm{XRD}(E_\nu)$. Here, $c_\mathrm{std} = \SI{1.63}{\pm}$ and \SI{1.80}{\pm} for simulated and experimental curves, respectively. This is the consequence of a slightly larger broadening of the experimental $I_\mathrm{XRD}$ curve due to the presence of an initial strain profile $\overline{\eta} \neq 0$, while $\overline{\eta} = 0$ for the simulated $I_\mathrm{XRD}$ curve.}
\centering
\label{SI:fig:I_XRD_exp_sim}
\end{figure*}

From $\eta(z,t)$, we calculated $c(z,t) = c_0[\eta(z,t)+1]$ for each unit cell along $z$ and subsequently the diffraction profiles in the range $\SI{0}{\ps}<t<\SI{20}{\ps}$ (\ref{SI:fig:T_strain_XRD_sim_2_7_mJ}d). A selection of diffraction curves at different $t$ (\ref{SI:fig:T_strain_XRD_sim_2_7_mJ}e) are then compared to the experimental data (\ref{SI:fig:T_strain_XRD_sim_2_7_mJ}f). The main features of the experimental data, i.e., the shift of the diffraction peak to smaller photon energies, the relative increase/decrease of the diffraction intensity on the low/high photon energy side, the larger broadening and smaller peak diffraction intensity as $t$ increases, are all well reproduced by our simulations. This strongly corroborates the validity of the model employed to describe our data. The initial shift to higher photon energy for $t < \SI{4}{\ps}$ is clearly shown by the increase of spectral weight around \SI{1530}{\eV} in \ref{SI:fig:T_strain_XRD_sim_2_7_mJ}d. Conversely, for $t > \SI{4}{\ps}$, the average positive strain explains the peak shift to lower photon energy, and the strain gradient is the reason for the change in spectral weight from the high to the low energy side. At the same time, a larger strain gradient leads to broader diffraction curves, and given the conservation of the total area under the curve, to a smaller peak intensity. Qualitatively, the simulated diffraction curves show sharper oscillations, related to the small film thickness, as compared to the experimental ones (\ref{SI:fig:T_strain_XRD_sim_2_7_mJ}e-f). This is the consequence of a slightly larger broadening of the experimental $I_\mathrm{XRD}$ curves that we assign to an initial strain profile $\eta(z,t=\SI{0}{ps}) \neq 0$, while we assume $\eta(z,t=\SI{0}{ps}) = 0$ throughout the sample for the simulated $I_\mathrm{XRD}$ curves (\ref{SI:fig:I_XRD_exp_sim}). Finally, the discussion above regarding the incident pump fluence $F_\mathrm{in} = \SI{2.7}{\mJ\per\square\cm}$ applies also to the lower fluence $F_\mathrm{in} = \SI{1.4}{\mJ\per\square\cm}$. Also in the latter case, we find an excellent agreement between experiments and simulations (\ref{SI:fig:T_strain_XRD_sim_1_6_mJ}e-f).

\clearpage

\clearpage

\clearpage

\clearpage

\section{Estimation of the bandgap decrease}\label{si:note:estimation_bandgap}

To estimate the largest bandgap decrease, first we determine the photoinduced electronic pressure as $\Delta P = \gamma_e C_e \Delta T_e \approx \SI{1.7e9}{\pascal}$ \cite{ruello_physical_2015}, where $\gamma_e \approx 2.6$ \cite{Choithrani2014StructuralEA} is the Grüneisen parameter, $C_e$ is the electronic heat capacity (\ref{SI:tab:constants}), and $\Delta T_e = \SI{1.56e4}{\kelvin}$ is the largest increase in electronic temperature of the BTO film averaged over $d_\mathrm{BTO}$ (\ref{si:fig:Temperature_profile}). Second, we express the bandgap variation as $\Delta E_g = \frac{\partial E_g}{\partial P} \Delta P \approx \SI{-3.2}{\milli\eV}$, where $\frac{\partial E_g}{\partial P} = \SI{-3.0(5)e-31}{\joule\per\pascal}$ (\ref{si:tab:fit_results}).

\section{Estimation of the electronic contribution to the ferroelectric polarization $P_s$}\label{si:note:estimation_electronic_contribution}

To quantify the relative contribution of structural ($\Delta P_s^\eta / P_s$) and electronic ($\Delta P_s^{n_e} / P_{s,0}$) changes to the total polarization variation out of equilibrium $\Delta P_s / P_{s,0} = (\Delta P_s^\eta + \Delta P_s^{n_e}) / P_{s,0} = \Delta P_s^\eta / P_{s,0} (1 + C_{n_e})$, we first estimate  $\Delta P_s^\eta / P_{s,0}$, and then derive the relative magnitude of the electronic contribution $C_{n_e}=(\Delta P_s^{n_e}/P_{s,0}) / (\Delta P_s^\eta / P_{s,0})$. From polarization-electric field hysteresis loops on differently strained BTO thin films grown on a GSO substrate \cite{pesquera_beyond_2020}, the relative change of the polarization as a function of the strain was determined to be $\Delta P_s^\eta/\Delta\eta = \SI{15}{\micro\coulomb\per\square\cm\per\%}$. In our data, at short time delays ($t < \SI{4}{\ps}$) the strain experiences a marginal change $\Delta \overline{\eta} < 0.02 \%$ and $\Delta P_s$ is clearly dominated by the presence of photoexcited carriers in the conduction band (Figure \ref{fig:Figure_3_shg_refl}e-f). At larger time delays ($t \approx \SI{20}{\ps}$), $\overline{\eta}$ reaches the saturation value of $0.34 \%$, which would correspond to an increase in polarization of $\Delta P_s^\eta = \SI{5.1}{\micro\coulomb\per\square\cm}$ ($\Delta P_s^\eta/P_{s,0} \approx 25 \%$), assuming a polarization at equilibrium of $P_{s,0} \approx \SI{20}{\micro\coulomb\per\square\cm}$ as found for BTO/SRO/GSO \cite{pesquera_beyond_2020}. In contrast, we measure an overall polarization change of $\Delta P_s/P_{s,0} \approx -2.5 \%$ (Figure \ref{fig:Figure_3_shg_refl}e). As a result, it follows that the electronic contribution $C_{n_e}$, responsible for the reduced polarization, has $\approx 10 \%$ larger magnitude than the structural one even at \SI{20}{\ps} ($C_{n_e} \approx -1.1$).

\clearpage

\bibliography{references}

\end{document}